\documentclass[twocolumn]{aastex631}
\usepackage{amsmath}
\usepackage{graphicx}
\usepackage{placeins}
\usepackage{float}
\usepackage{txfonts}
\usepackage{hyperref}

\usepackage{lipsum}

\newcommand\blfootnote[1]{%
  \begingroup
  \renewcommand\thefootnote{}\footnote{#1}%
  \addtocounter{footnote}{-1}%
  \endgroup
}

\clearpage

\newcommand{\Co}{$\rm ^{56}Co$}
\newcommand{\Fe}{$\rm ^{56}Fe$}

\begin{document}

\title{A Sensitive Search for Supernova Emission Associated with the Extremely Energetic and Nearby GRB 221009A}

\correspondingauthor{Gokul P. Srinivasaragavan}\email{gsriniv2@umd.edu}
\author[0000-0002-6428-2700]{Gokul P. Srinivasaragavan}
\affiliation{Department of Astronomy, University of Maryland, College Park, MD 20742, USA}
\affiliation{Joint Space-Science Institute, University of Maryland, College Park, MD 20742, USA}
 \affiliation{Astrophysics Science Division, NASA Goddard Space Flight Center, 8800 Greenbelt Rd, Greenbelt, MD 20771, USA}
\author[0000-0002-9700-0036]{Brendan O'Connor}
    \affiliation{Department of Physics, The George Washington University, Washington, DC 20052, USA}
    \affiliation{Astronomy, Physics and Statistics Institute of Sciences (APSIS), The George Washington University, Washington, DC 20052, USA}
   \affiliation{Department of Astronomy, University of Maryland, College Park, MD 20742, USA}
    \affiliation{Astrophysics Science Division, NASA Goddard Space Flight Center, 8800 Greenbelt Rd, Greenbelt, MD 20771, USA}
\author[0000-0003-1673-970X]{S. Bradley Cenko}
\affiliation{Astrophysics Science Division, NASA Goddard Space Flight Center, 8800 Greenbelt Rd, Greenbelt, MD 20771, USA}
\affiliation{Joint Space-Science Institute, University of Maryland, College Park, MD 20742, USA}

\author[0000-0001-6157-6722]{Alexander J.~Dittmann}
\affiliation{Department of Astronomy, University of Maryland, College Park, MD 20742, USA}
\affiliation{Joint Space-Science Institute, University of Maryland, College Park, MD 20742, USA}

\author[0000-0002-2898-6532]{Sheng Yang}
\affiliation{Henan Academy of Sciences, Zhengzhou 450046, Henan, China}
\affiliation{Department of Astronomy, The Oskar Klein Center, Stockholm University, AlbaNova, 10691 Stockholm, Sweden}

\author[0000-0003-1546-6615]{Jesper Sollerman}
\affiliation{Department of Astronomy, The Oskar Klein Center, Stockholm University, AlbaNova, 10691 Stockholm, Sweden}

\author[0000-0003-3533-7183]{G.C. Anupama}
\affiliation{Indian Institute of Astrophysics, 2nd Block 100 Feet Rd, Koramangala Bangalore, 560 034, India}

\author[0000-0002-3927-5402]{Sudhanshu Barway}
\affiliation{Indian Institute of Astrophysics, 2nd Block 100 Feet Rd, Koramangala Bangalore, 560 034, India}

\author[0000-0002-6112-7609]{Varun Bhalerao}
\affiliation{Department of Physics, Indian Institute of Technology Bombay, Powai, 400 076, India}

\author[0000-0003-0871-4641]{Harsh Kumar}
\affiliation{Department of Physics, Indian Institute of Technology Bombay, Powai, 400 076, India}

\author[0000-0002-7942-8477]{Vishwajeet Swain}
\affiliation{Department of Physics, Indian Institute of Technology Bombay, Powai, 400 076, India}

\author[0000-0002-5698-8703]{Erica Hammerstein}
\affiliation{Department of Astronomy, University of Maryland, College Park, MD 20742, USA}
\affiliation{Astrophysics Science Division, NASA Goddard Space Flight Center, 8800 Greenbelt Rd, Greenbelt, MD 20771, USA}
\affiliation{Center for Research and Exploration in Space Science and Technology, NASA/GSFC, Greenbelt, MD 20771, USA}

\author[0000-0002-3097-942X]{Isiah Holt}
\affiliation{Department of Astronomy, University of Maryland, College Park, MD 20742, USA}
\affiliation{Joint Space-Science Institute, University of Maryland, College Park, MD 20742, USA}
 \affiliation{Astrophysics Science Division, NASA Goddard Space Flight Center, 8800 Greenbelt Rd, Greenbelt, MD 20771, USA}

\author[0000-0003-3768-7515]{Shreya Anand}
\affiliation{Division of Physics, Mathematics and Astronomy, California Institute of Technology, Pasadena, CA 91125, USA}

\author[0000-0002-8977-1498]{Igor Andreoni$^\dag$}
\blfootnote{$^\dag$Gehrels Fellow}
\affiliation{Joint Space-Science Institute, University of Maryland, College Park, MD 20742, USA}
\affiliation{Department of Astronomy, University of Maryland, College Park, MD 20742, USA}
 \affiliation{Astrophysics Science Division, NASA Goddard Space Flight Center, 8800 Greenbelt Rd, Greenbelt, MD 20771, USA}

\author[0000-0002-8262-2924]{Michael~W.~Coughlin}
\affiliation{School of Physics and Astronomy, University of Minnesota,
Minneapolis, Minnesota 55455, USA}

\author[0000-0001-6849-1270]{Simone Dichiara}
\affiliation{Department of Astronomy and Astrophysics, The Pennsylvania State University, 525 Davey Lab, University Park, PA 16802, USA}
\author[0000-0002-3653-5598]{Avishay Gal-Yam}
    \affiliation{Department of Particle Physics and Astrophysics, Weizmann Institute of Science, 76100 Rehovot, Israel}
 
\author[0000-0002-2666-728X]{M. Coleman Miller}
\affiliation{Department of Astronomy, University of Maryland, College Park, MD 20742, USA}
\affiliation{Joint Space-Science Institute, University of Maryland, College Park, MD 20742, USA}

\author[0000-0001-9226-4043]{Jaime Soon}
\affiliation{Research School of Astronomy and Astrophysics, Australian National University, Cotter Rd, Weston Creek ACT 2611, Australia}

\author[0000-0002-4622-796X]{Roberto Soria}
\affiliation{College of Astronomy and Space Sciences, University of the Chinese Academy of Sciences, Beijing 100049, China}
\affiliation{INAF - Osservatorio Astrofisico di Torino, Strada Osservatorio 20, I-10025 Pino Torinese, Italy}
\affiliation{Sydney Institute for Astronomy, School of Physics A28, The University of Sydney, NSW 2006, Australia}

\author[0000-0002-3774-1270]{Joseph Durbak}
\affiliation{Department of Physics, University of Maryland, College Park, MD 20742, USA}
\affiliation{Astrophysics Science Division, NASA Goddard Space Flight Center, 8800 Greenbelt Rd, Greenbelt, MD 20771, USA}
\affiliation{Center for Research and Exploration in Space Science and Technology, NASA/GSFC, Greenbelt, MD 20771, USA}

\author[0000-0002-8094-6108]{James H. Gillanders}
\affiliation{Department of Physics, University of Rome ``Tor Vergata’’, via della Ricerca Scientifica 1, I-00133 Rome, Italy}

\author[0000-0003-2714-0487]{Sibasish Laha}
\affiliation{Center for Space Science and Technology, University of Maryland Baltimore County, 1000 Hilltop Circle, Baltimore, MD 21250, USA.}
 \affiliation{Astrophysics Science Division, NASA Goddard Space Flight Center, 8800 Greenbelt Rd, Greenbelt, MD 20771, USA}
\affiliation{Center for Research and Exploration in Space Science and Technology, NASA/GSFC, Greenbelt, MD 20771, USA}

\author[0000-0002-2894-6936]{Anna M. Moore}
    \affiliation{Australian National University, Research School of Astronomy and Astrophysics, Mount Stromlo Observatory, Cotter Road, Weston Creek 2611, Australia }

\author[0000-0003-2132-3610]{Fabio Ragosta}
\affiliation{INAF, Osservatorio Astronomico di Roma, via Frascati 33, I-00078
Monte Porzio Catone (RM), Italy
}

\author[0000-0002-1869-7817]{Eleonora Troja}
\affiliation{Department of Physics, University of Rome ``Tor Vergata’’, via della Ricerca Scientifica 1, I-00133 Rome, Italy}
\affiliation{INAF - Istituto di Astrofisica e Planetologia Spaziali, via Fosso del Cavaliere 100, 00133 Rome, Italy}
 
\begin{abstract}
We report observations of the optical counterpart of the long gamma-ray burst (LGRB) GRB 221009A. Due to the extreme rarity of being both nearby ($z = 0.151$) and highly energetic ($E_{\gamma,\mathrm{iso}} \geq 10^{54}$\,erg), GRB 221009A offers a unique opportunity to probe the connection between massive star core collapse and relativistic jet formation across a very broad range of $\gamma$-ray properties. Adopting a phenomenological power-law model for the afterglow and host galaxy estimates from high-resolution Hubble Space Telescope imaging, we use Bayesian model comparison techniques to determine the likelihood of an associated SN contributing excess flux to the optical light curve. Though not conclusive, we find moderate evidence ($K_{\rm{Bayes}}=10^{1.2}$) for the presence of an additional component arising from an associated supernova, SN 2022xiw, and find that it must be substantially fainter ($< 67 \%$ as bright at the 99\% confidence interval) than SN 1998bw. Given the large and uncertain line-of-sight extinction, we attempt to constrain the supernova parameters ($M_{\mathrm{Ni}}$, $M_{\mathrm{ej}}$, and $E_{\mathrm{KE}}$) under several different assumptions with respect to the host galaxy's extinction. We find properties that are broadly consistent with previous GRB-associated SNe: $M_{\rm{Ni}}=0.05$ -- $0.25 \, \rm{M_\odot}$, $M_{\rm{ej}}=3.5$ -- $11.1 \, \rm{M_\odot}$, and $E_{\rm{KE}} = (1.6$ -- $5.2) \times 10^{52} \, \rm{erg}$. We note that these properties are weakly constrained due to the faintness of the supernova with respect to the afterglow and host emission, but we do find a robust upper limit on the $M_{\rm{Ni}}$ of $M_{\rm{Ni}}<0.36\, \rm{M_\odot}$. Given the tremendous range in isotropic gamma-ray energy release exhibited by GRBs (7 orders of magnitude), the SN emission appears to be decoupled from the central engine in these systems. 
\end{abstract}

\section{Introduction}
\label{sec:intro}
Over the past two and a half decades, a link has been established between long-duration gamma-ray bursts (LGRBs) and core-collapse supernovae (CCSNe; \citealt{Woosley2006}). Over two dozen LGRBs have been associated with CCSNe, either indirectly (e.g., through late-time ``bumps'' in their optical afterglow light curves), or directly through tell-tale spectroscopic signatures \citep{cano2017}. All of these SNe are of the Ic-BL type \citep{Filippenko1997}: they lack H and He lines in their optical spectra, and possess broad lines indicative of higher ejecta velocities than seen in normal Type Ic SNe.

Despite this progress, a number of key open questions regarding the nature of the GRB-SN connection remain. One of the foremost of these is understanding why a much smaller fraction of Type Ic-BL SNe have associated LGRBs than the converse. A number of studies \citep{Soderberg+2006,Corsi2016,Corsi2022} have shown that this dichotomy cannot be explained solely by viewing angle effects, and that relativistic ejecta are not ubiquitous to SNe Ic-BL ($<19\%$ of Ic-BL events are SN 1998bw-like, the prototypical SN associated with a GRB; \citealt{Corsi2022}). Therefore, there are intrinsic differences in the explosion mechanisms and/or environments between jet-powered Type Ic-BL SNe and normal Ic-BL events, and understanding this dichotomy can provide important insights into stellar evolution and the landscape of stellar explosions.

The observed population of LGRBs is comprised predominantly of cosmological ($z \gtrsim 1$) events, with $E_{\gamma,\mathrm{iso}}$ between $10^{50}$ to $10^{54}$\,erg (e.g., \citealt{Jakobsson+2006}). On the other hand, the majority of GRBs that have associated spectroscopically confirmed SNe (GRB-SN) are low-luminosity events, with isotropic equivalent energies ($E_{\gamma,\mathrm{iso}}$) between $10^{48}$ to $10^{50}$ erg \citep{cano2017}. This is the natural consequence of low-luminosity GRBs dominating the population of events nearby enough for spectroscopic investigations ($z \lesssim 0.3$), even with large-aperture optical facilities. Here we attempt to characterize the supernova associated with a rare energetic LGRB discovered in the nearby universe.

GRB 221009A [$\alpha$ (J2000)= 19$^{\mathrm{h}}$13$^{\mathrm{m}}$03.50$^{\mathrm{s}}$,
$\delta$ (J2000) = +19$^{\circ}$46$\arcmin$24$\farcs$23; \citealt{GCN.32757}] was discovered by the Burst Alert Telescope (BAT; \citealt{Barthelmy2005}) on the Neil Gehrels Swift Observatory (Swift; \citealt{gehrels_2004}), though initially classified as a potential Galactic transient \citep{GCN32632}. Subsequently, the Fermi Gamma-Ray Burst Monitor (GBM; \citealt{Meegan+2009}) reported an extremely bright LGRB detected $\approx$ 55\,min earlier\footnote{At 13:16:59.00 UTC on 9 Oct 2022, which we establish hereafter as $\rm{T_0}$.} consistent with this localization \citep{GBMGCN}. Further observations revealed that the BAT triggered on the bright X-ray afterglow of GRB 221009A, a first in the nearly 18 years of Swift operations \citep{Williams+2023}.

The unprecedented brightness led to extensive follow-up across the electromagnetic spectrum (e.g., \citealt{Fulton2023,Williams+2023,Laskar+2023,OConnor+2023,Kann+2023}). Spectroscopy of the optical afterglow led to a redshift measurement of $z = 0.151$ \citep{XshooterGCN,GCN.32686,Malesani2023}. Its associated gamma-ray isotropic energy release is well in excess of $10^{54} \, \rm{erg}$, making GRB 221009A an extremely rare example of a highly energetic LGRB nearby enough to search for an associated SN Ic-BL.

In this \textit{Letter}, we present optical observations that display a late-time flattening in the afterglow decay of GRB 221009A, and we investigate how these measurements can constrain the possible associated SN (SN 2022xiw; \citealt{TNS}). In \S \ref{Opobservations} we report the observations of the optical afterglow of GRB 221009A; in \S\ref{analysis} we analyze the observations, and statistically compare models with and without a supernova component; in \S\ref{LCModeling} we constrain physical parameters of SN 2022xiw; in \S \ref{Comparison} we place GRB 221009A/SN 2022xiw in the context of the GRB-SN sample in literature; and in \S \ref{Conclusion} we summarize our conclusions. In the final stages of manuscript preparation, studies reporting conflicting results on the existence of SN emission were posted on the arXiv \citep{Fulton2023,Shrestha+2023,Kann+2023,Levan2023} -- where relevant we highlight differences in our approach and contrasts in our results.


\section{Observations} 
\label{Opobservations}
The main telescopes used for this work are the GROWTH-India Telescope (GIT), Lowell Discovery Telescope (LDT), and the Gemini-South Telescope. The results of the GIT, LDT, and Gemini-South observations are presented in Table~\ref{Table1}. To provide additional coverage at early times and help constrain the afterglow behavior, we have also considered optical afterglow measurements from the Liverpool Telescope (LT) reported in \citet{Laskar+2023}, and supplemented these with preliminary results reported in the GCN circulars \citep{GCN.32769, GCN.32818, GCN.32684, GCN.32743, GCN.32752, GCN.32652, GCN.32755, GCN.32646, GCN.32804, GCN.32678, GCN.32811, GCN.32758, GCN.32803, GCN32765hostz, GCN.32670,GCN.32662, GCN.32852, GCN.32795, GCN.32809, GCN.32753,GCN.32771, GCN.32709}. The GCN photometry is provided in tabular form in the Appendix. All times used in this work are in the observer frame. Below we describe the data reduction processes for each of the telescopes we use. 

\begin{longtable}{ccccc}
\hline
\hline
$t_{\rm{obs}} - T_0$ (days)  & Telescope & Filter & AB mag & Uncertainty \\
\hline
       0.12835635 & GIT & g$'$ & 17.66 & 0.08 \\ 
        0.13105835 & GIT & r$'$ & 16.16 & 0.07 \\ 
        0.13366335 & GIT & i$'$ & 15.17 & 0.03 \\ 
        0.13633535 & GIT & z$'$ & 14.50 & 0.04 \\ 
        0.14391035 & GIT & g$'$ & 17.69 & 0.07 \\ 
        0.14653335 & GIT & r$'$ & 16.26 & 0.05 \\ 
        0.14922235 & GIT & i$'$ & 15.28 & 0.03 \\ 
        0.15183035 & GIT & z$'$ & 14.57 & 0.05 \\ 
        0.15955335 & GIT & g$'$ & 17.99 & 0.10 \\ 
        0.16225435 & GIT & r$'$ & 16.32 & 0.05 \\ 
        0.16486235 & GIT & i$'$ & 15.34 & 0.04 \\ 
        0.16755835 & GIT & z$'$ & 14.64 & 0.06 \\ 
        0.17138035 & GIT & g$'$ & 17.94 & 0.11 \\ 
        0.17407535 & GIT & r$'$ & 16.40 & 0.05 \\ 
        0.17668335 & GIT & i$'$ & 15.41 & 0.04 \\ 
        0.17939935 & GIT & z$'$ & 14.71 & 0.07 \\ 
        0.18711135 & GIT & g$'$ & 18.11 & 0.16 \\ 
        0.18971935 & GIT & r$'$ & 16.49 & 0.05 \\ 
        0.19239635 & GIT & i$'$ & 15.47 & 0.05 \\ 
        0.19499735 & GIT & z$'$ & 14.72 & 0.06 \\ 
        1.03149435 & GIT & g$'$ & 20.21 & 0.23 \\ 
        1.03551335 & GIT & z$'$ & 16.75 & 0.05 \\ 
        1.03843935 & GIT & z$'$ & 16.73 & 0.06 \\ 
        1.04287735 & GIT & i$'$ & 17.48 & 0.04 \\ 
        1.04447835 & GIT & i$'$ & 17.49 & 0.03 \\ 
        1.04607935 & GIT & i$'$ & 17.52 & 0.04 \\ 
        1.05022335 & GIT & r$'$ & 18.58 & 0.06 \\ 
        1.05371835 & GIT & r$'$ & 18.55 & 0.05 \\ 
        1.08636135 & GIT & g$'$ & 20.50 & 0.16 \\ 
        1.09184435 & GIT & z$'$ & 16.90 & 0.15 \\ 
        1.10147035 & GIT & r$'$ & 18.66 & 0.08 \\ 
        1.14527035 & GIT & z$'$ & 17.06 & 0.08 \\ 
        1.14817435 & GIT & z$'$ & 17.01 & 0.15 \\ 
        2.11955235 & GIT & r$'$ & 19.77 & 0.14 \\ 
        2.14304035 & GIT & z$'$ & 17.99 & 0.15 \\ 
        3.00313235 & GIT & i$'$ & 19.20 & 0.14 \\ 
        3.00693735 & GIT & r$'$ & 20.41 & 0.15 \\ 
        3.00697835 & GIT & z$'$ & 18.57 & 0.16 \\ 
        4.02688035 & GIT & i$'$ & 19.68 & 0.14 \\ 
        4.03068335 & GIT & r$'$ & 20.83 & 0.16 \\ 
        4.03069335 & GIT & z$'$ & 18.82 & 0.17 \\ 
        5.07656435 & GIT & r$'$ & 20.91 & 0.21 \\ 
        5.07656635 & GIT & i$'$ & 19.85 & 0.19 \\ 
        5.08403035 & GIT & z$'$ & 18.93 & 0.17 \\ 
        6.09907835 & GIT & i$'$ & 20.43 & 0.18 \\ 
        6.10467335 & GIT & z$'$ & 19.41 & 0.19 \\ 
        3.6 & LDT & g$'$ & 22.06 & 0.04 \\
        3.6 & LDT & r$'$ & 20.44 & 0.04 \\ 
        3.6 & LDT & i$'$ & 19.37 & 0.01 \\
        3.6 & LDT & z$'$ & 18.67 & 0.01 \\
        9.5 & LDT & r$'$  & 21.5 & 0.08 \\ 
        9.5 & LDT & i$'$ & 20.75 & 0.05 \\ 
        9.5 & LDT & z$'$ & 20.01 & 0.03 \\
        18.5 & LDT & g$'$ & 24.71 & 0.15 \\
        18.5 & LDT & r$'$ & 22.71 & 0.06 \\ 
        18.5 & LDT & i$'$ & 21.83 & 0.15 \\ 
        18.5 & LDT & z$'$ & 20.97 & 0.05 \\
        21.5 & LDT & r$'$ & 22.91 & 0.06 \\ 
        21.5 & LDT & i$'$ & 21.82 & 0.02 \\
        21.5 & LDT & z$'$ & 21.37 & 0.04 \\ 
        28.5 & LDT & r$'$ & 23.55 & 0.15 \\ 
        28.5 & LDT & i$'$ & 22.07 & 0.09 \\ 
        28.5 & LDT & z$'$ & 21.45 & 0.07 \\
        36.53 & LDT & r$'$ & 23.64 & 0.11 \\ 
        36.53 & LDT & i$'$ & 22.45 & 0.05 \\ 
        36.53 & LDT & z$'$ & 21.82 & 0.07 \\ 
        52.53 & LDT & i$'$ & 22.73 & 0.15\\
        4.4 & Gemini-South & i$'$ & 19.78 & 0.02 \\ 
        17.4 & Gemini-South & i$'$ & 21.71 & 0.05 \\
        \hline
 \caption{Optical photometry and associated 1$\sigma$ errors of GRB 221009A, which includes contributions from its afterglow, host galaxy, and associated SN 2022xiw. All times are in the observer frame. The magnitudes are not corrected for Galactic extinction.}
 \label{Table1}
\end{longtable}

\subsection{GIT}
We used the GIT located at the Indian Astronomical Observatory (IAO), Hanle-Ladakh, to acquire observations of the GRB 221009A optical afterglow \citep{2022AJ....164...90K}. The source was observed in Sloan g$'$, r$'$, i$'$ and z$'$ bands. Data were downloaded and processed in real-time by the GIT data reduction pipeline. We used individual exposures for photometry in the early stages when the afterglow was bright. Later, we stacked images with \texttt{SWarp} \citep{2010ascl.soft10068B} to increase the S$/$N ratio of detections. 

\begin{figure*}
    \centering
    \includegraphics[width=1\linewidth]{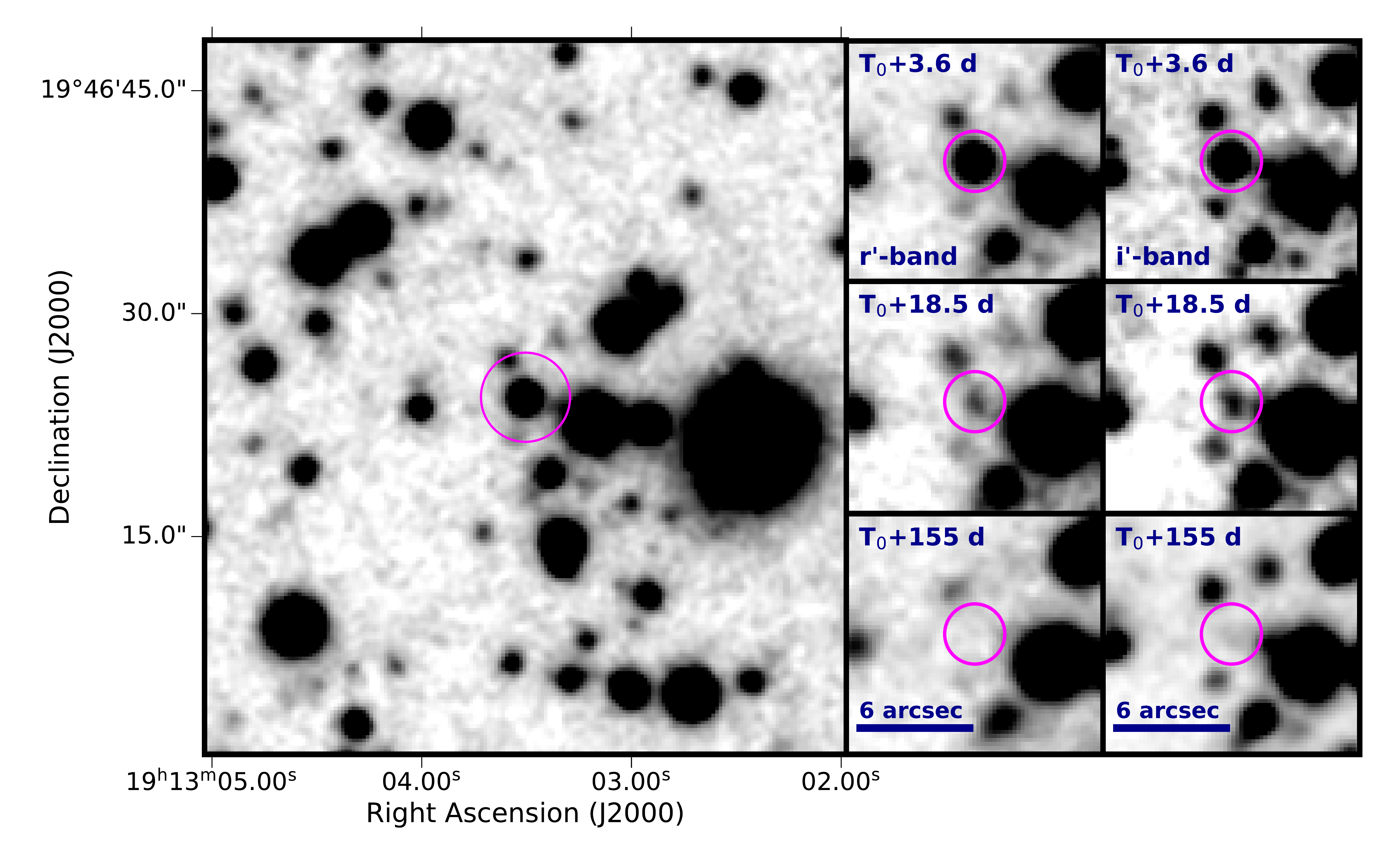}
    \caption{LDT images of the optical afterglow of GRB 221009A. All of the panels show the position of GRB 221009A circled in pink. The left, large panel shows the wider field of view of GRB 221009A, in r$'$ band 3.6 days after $\rm{T_0}$. The right panels show the evolution over time of the optical afterglow, in both r$'$ and i$'$ bands. The image at 155 days is devoid of afterglow and SN contribution, and the host galaxy is faintly seen in i$'$ band. The images have been smoothed for display purposes.
    }
     \label{LDTobs}
\end{figure*}

The data were reduced in a standard manner using the GIT pipeline \citep{2022MNRAS.516.4517K}. All images were pre-processed by subtracting bias, flat-fielding, and cosmic-ray removal via the \texttt{Astro-SCRAPPY} \citep{2019ascl.soft07032M} package. Astrometry was performed on the resulting images using the offline \texttt{solve-field} astrometry engine. Sources were detected using \texttt{SExtractor} \citep{1996AAS..117..393B} and were crossed-matched against the Pan-STARRS1 DR1 catalog (PS1; \citealt{PS1}) through \texttt{VizieR} to obtain the zero-point. Finally, the pipeline performed point spread function (PSF) fit photometry to obtain the magnitudes of the GRB 221009A afterglow (Table~\ref{Table1}).

\subsection{LDT}
We also observed GRB 221009A in r$^{\prime}$, i$^{\prime}$, and z$^{\prime}$ with the 4.3m Large Monolithic Imager (LMI) on the LDT through an approved target-of-opportunity (ToO) program. We reduced the images using a custom \texttt{Python}-based image analysis pipeline \citep{Toy2016}, that can perform data reduction, astrometry, registration, source extraction and PSF photometry using \texttt{SExtractor}, which was calibrated using point-sources from the PS1 catalog (Table~\ref{Table1}). These observations were also reported in \citet{GCN32739,GCN32799,OConnor+2023}. Figure \ref{LDTobs} shows both the wider field of view of GRB 221009A's position on the sky, as well as the individual evolution of its flux over time in both filters.

\subsection{Gemini-South}
Two additional publicly available i$'$-band observations obtained with GMOS-S mounted on the 8.1m Gemini-South Telescope (PI: Rastinejad, O'Connor; \citealt{GCN32749}) were also analyzed. The data were reduced using \texttt{DRAGONS}\footnote{\url{https://dragons.readthedocs.io/}} \citep{Labrie2019} to align and stack individual frames. PSF photometry was calibrated using nearby point-sources in the PS1 catalog (Table~\ref{Table1}).


\section{Analysis and Model Selection}
\label{analysis}

\begin{figure*}
    \centering
    \includegraphics[width=0.75\linewidth]{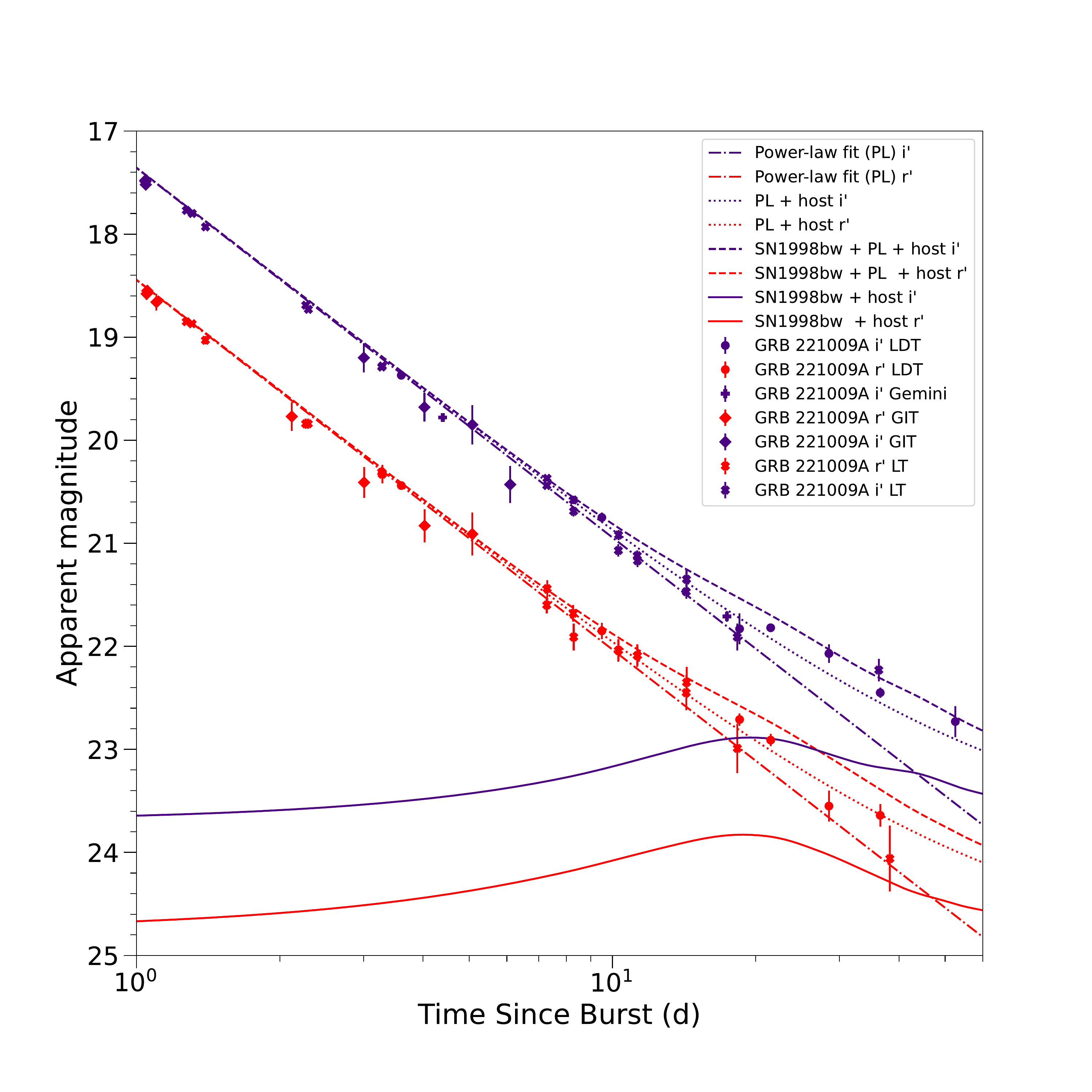}
    \caption{Observed r$'$- and i$'$-band photometry of GRB 221009A after $T_0 + 1 \, \rm{d}$, along with the best-fit optical afterglow model corresponding to a power-law decay index of $\alpha = 1.434$, with the addition of the host galaxy emission. The LCs for a SN 1998bw-like source in r$'$ and i$'$ band, redshifted to $z = 0.151$ and reddened according to the Galactic and host galaxy extinction of GRB 221009A, are also shown. }
    \label{afterglowphotometry}
\end{figure*}

\subsection{Previous Broadband Modeling Results}
\label{previousbroad}
We draw upon three results from previous studies of the broadband afterglow of GRB 221009A in order to inform our approach to characterizing the associated SN emission. First, fits to the broadband spectral energy distribution (SED) revealed that the frequency range from optical to hard X-rays is not well fit by a single power-law (i.e., $f_{\nu} \propto \nu^{-\beta}$). Instead, a change in spectral slope, physically attributed to the synchrotron cooling frequency ($\nu_{c}$), is inferred around the X-ray bandpass \citep{Williams+2023,OConnor+2023}.  
As a result, in standard synchrotron afterglow theory, we do not expect the optical afterglow to decay with the same power-law index as the X-rays (e.g., \citealt{Sari1998}). Thus, to remove the afterglow contribution we must model the optical decay separately from the X-rays (c.f., \citealt{Fulton2023} and the first approach in \citealt{Shrestha+2023}). 

Second, in addition to significant extinction due to dust in the Milky Way galaxy, broadband SED fits indicate the possible existence of absorption beyond the nominal value of $E(B-V) = 1.30$\,mag reported in \citet{Schlafly2011}. Given the low redshift, it is difficult to disentangle a larger than expected Galactic extinction (e.g., due to small-scale variations in Galactic dust), or extinction in the GRB host galaxy. Existing works differ on the significance of this extinction component, with inferred values ranging from $E(B-V) = 1.30$ mag (i.e., no additional extinction; \citealt{OConnor+2023}) to $E(B-V) = 1.80$ mag \citep{Fulton2023,Williams+2023}. Given these uncertainties, we consider the implications of differing host extinction levels throughout this work.

Finally we correct for underlying host galaxy light in our analysis, derived by \citet{Levan2023} through \texttt{GALFITM} modeling of late-time HST observations. We use their measurements of $\rm{F625W} = 24.88 \pm 0.08$ mag, and  $\rm{F775W} = 23.80 \pm 0.14$ mag, which approximately correspond to the r$'$ and i$'$ bands. Previous analyses \citep{Fulton2023,Shrestha+2023,Kann+2023} did not incorporate the host contribution explicitly in their analysis (though Pan-STARRS and DECam photometry in \citealt{Fulton2023}, and DECam photometry in \citealt{Shrestha+2023} is subjected to template subtraction, which negates the host contribution to a degree).
\subsection{Optical Afterglow Modeling}
\label{oag}
Assuming the optical afterglow is powered by synchroton emission from the forward shock \citep{Meszaros1997}, we fit the early-time light curve with power-law models ($f_{\nu} \propto t^{-\alpha}$) to attempt to isolate the contribution from the supernova. Prior to $T_{0} + 1$\,d, the optical data display a shallow initial slope with $\alpha=0.88\pm0.05$ \citep{OConnor+2023, Kann+2023}. Beyond this time there is a clear steepening in the decay, and a single power-law does not provide a good fit to all optical data \citep{GCN.32755}.

Since a SN will contribute negligibly (compared to the bright afterglow here) in the first days post explosion, we perform a power-law fit of all g$'$, r$'$, i$'$, and z$'$ data between 1 and 6 days post-explosion. We find a best fit decay index of $\alpha = 1.434 \pm 0.004$. As found by other authors, this index is appreciably shallower than the X-ray decay found at this time \citep{Williams+2023,Laskar+2023,OConnor+2023}. Figure \ref{afterglowphotometry} shows the best-fit power-law derived, along with the addition of the host galaxy emission to that power law.

The late-time behavior shows possible deviations from this power-law, even after the host galaxy emission is accounted for. Though it is possible that central-engine activity can cause re-brightening of the optical afterglow in excess of what is expected from a power-law decay, this usually occurs directly after the prompt emission in the early-time evolution of the GRB \citep{Kann2007, Oates2009}. Because the brightening occurs weeks after the initial prompt emission (see Figure~\ref{LDTobs}), we determine that there is a possibility that an associated SN is contributing excess flux to the optical emission, and further investigate this in \S \ref{ModelSelection}. 

\subsection{Model Selection}
\label{ModelSelection}
\begin{figure*}
    \centering
    \includegraphics[width=0.49\linewidth]{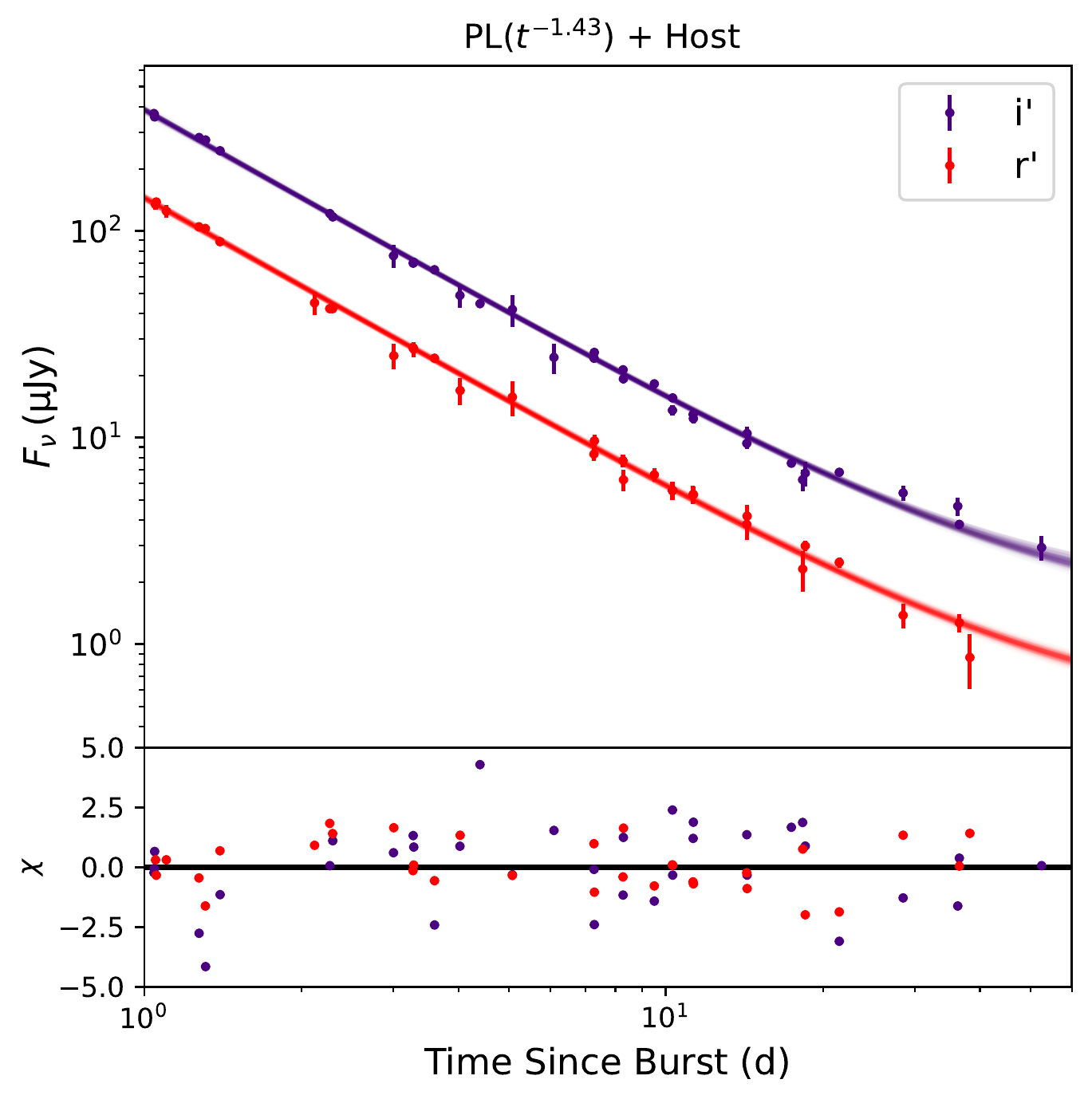}
    \includegraphics[width=0.49\linewidth]{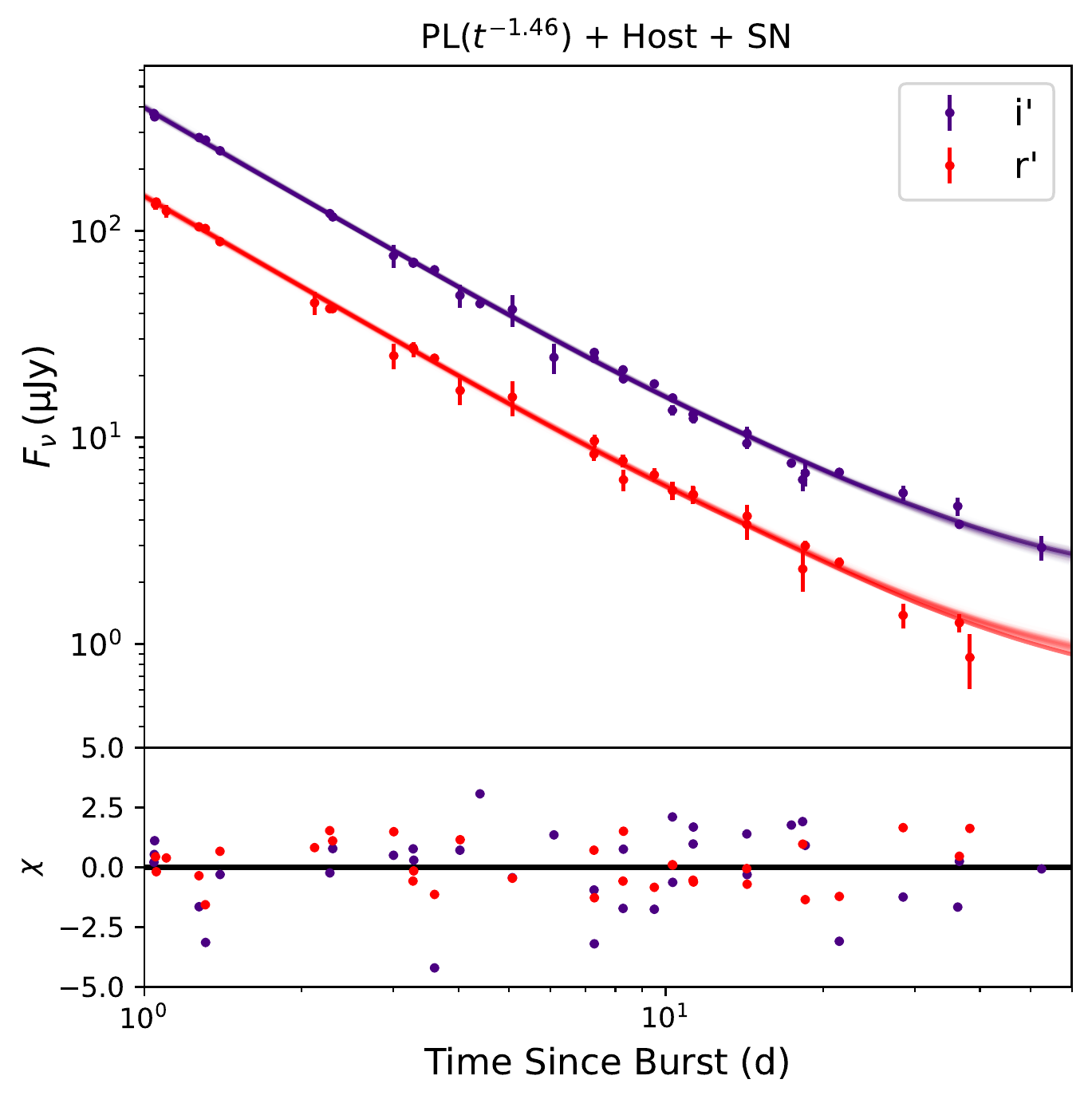}
    \caption{\textit{Left panel}: The best-fit afterglow+host model, along with its a posteriori possible models and $\chi$ values.  \textit{Right panel}: The best-fit SN model, along with its a posteriori models and $\chi$ values. The SN model is moderately but not conclusively favored, with a Bayes factor $K_{\rm{Bayes}} = 10^{1.2}$, and the best-fit parameters for each model are shown in Table \ref{modelselectiontable}.}
    \label{best-fitmodels}
\end{figure*}
In order to evaluate the statistical significance of the excess late-time emission, we perform a Bayesian model selection using the {\sc PyMultiNest} package \citep{Feroz09, Buchner+2014}. We consider two empirical models -- one where the optical emission is explained as the sum of the afterglow (i.e., synchrotron radiation) and a (constant) host galaxy emission, and one where an additional SN component is added as well. For each model, we calculate the Bayesian evidence, and use those the calculate the Bayes factor to see which is statistically preferred \citep{Trota08}. We limit the analysis to the r$'$ and i$'$ bands, as the large Galactic extinction makes g$'$ band uninformative, and z$'$ band SN templates are less available due to a lack of spectral coverage in that wavelength range.

For the afterglow+host model, we assume a single power-law decay in both bands at times after $T_{0}+1$\,d. While the index is fixed to be identical in both the r$'$ and i$'$ bands, we do not require its value be equal to that derived from 1--6\,d post-explosion (\S\ref{oag}); rather, we allow the index to vary to provide the best fit to the entire data set. The host galaxy emission is incorporated as a free parameter in the fit, using a Gaussian prior with the mean and standard deviation measured by HST \citep{Levan2023}. 

For the SN model, an additional component is added to the afterglow+host model to mimic SN behavior. We take the LC of SN 1998bw \citep{98bwpaper}, and (de)redden and K-correct it to match the relevant properties of GRB 221009A, using {\sc SNCosmo} \citep{SNCosmo}. Specifically we adopt $E(B-V)_{MW} = 1.30$\,mag \citep{Schlafly2011}, $E(B-V)_{host} = 0.3$\, mag, which is the most conservative value found in \citep{Williams+2023}, and apply the Milky Way extinction law from \citet{Cardelli1989} with $R_{V} = 3.1$. The resulting r$'$ and i$'$-band LCs for a SN 1998bw-like source at $z=0.151$ [and behind an extinction of $E(B-V)_{\mathrm{MW}} = 1.30$\,mag and $E(B-V)_{\mathrm{host}} = 0.3$\,mag] are shown in Figure~\ref{afterglowphotometry}.

The final SN model used in the model selection has two free parameters: a flux-stretching factor $k_{\mathrm{SN \, 1998bw}}$ and a time-stretching factor $s_{\rm{SN \, 1998bw}}$ (e.g., \citealt{Klose2019}). Both of these parameters have priors drawn from a bivariate normal distribution fit to values derived for previous GRB-SN \citep{cano2017}. We also allow the afterglow power-law decay index ($\alpha$) to be free with uniform priors, the flux constants of proportionality ($a_{\mathrm{AG}}$) to be free with uniform priors in log space, and add the host galaxy emission in both bands utilizing the same Gaussian priors. Therefore, the full SN model is:
\begin{equation}
 f_\nu({t_{\rm{obs}}}) = k_{\rm{SN}} (f_\nu^{\rm{SN 1998bw}}(t_{\rm{obs}}/s_{\rm{SN\, 1998bw}})) + a\\_{\rm{AG}} (t_{\rm{obs}})^{-\alpha} + f_{\nu}^{\rm{host}},
 \end{equation} 
where $ f_\nu^{\rm{SN\,1998bw}}(t_{\rm{obs}})$ is the flux seen of the SN at a time in the observer frame, $t_{\rm{obs}}$ is the time in the observer frame, and $f_{\nu}^{\rm{host}}$ is the flux contribution from the host galaxy. 

 Initially, we fit the two models concurrently to the r$'$ and i$'$-bands, while assuming the errors are the nominal values reported in Table \ref{Table1} and \citet{Laskar+2023} for the LT photometry. We calculate the Bayes factor ($K_{\rm{Bayes}}$) between the two models to determine the likelihood that a SN is contributing excess flux to the optical afterglow in addition to the host galaxy emission, and find $K_{\rm{Bayes}}=10^{4.0}$, which indicates that the SN model is strongly favored. However, when we calculate the $\chi^2$ statistic for each of the models, we find that $\chi^2_{\rm{PL+host}} = 120.83$ for 62 degrees of freedom and $\chi^2_{\rm{PL+host+SN}} = 112.62$ for 60 degrees of freedom. Though the $\Delta \chi^{2} = 8.21$ for 2 additional degrees of freedom is also indicative of a preference for the SN model, the $\chi^2$ statistics themselves indicate that neither model adequately fits the data. This is likely because in the initial fitting, we did not account for systematic uncertainties that arise from combining data from multiple telescopes in the full photometric data set, or S-corrections \citep{Stritzinger2002}.
 
 Therefore, we modify the fitting procedure to numerically optimize the likelihood function, where we assume that the reported errors under-estimate the true uncertainty. To account for this, we include an error-stretching factor ($\beta$) in the fitting procedure to represent the S-correction effect, and recalculate $K_{\rm{Bayes}}$. The log-likelihood function we minimize, with the addition of the error-stretching factor, is:
 \begin{equation}
    \mathrm{ln}\, p(y\, |\, m, s_n) =  -\frac{1}{2} \sum\limits_{n} \frac{(y_n-m_n)^2}{s_n^2} + \mathrm{ln}(2\pi s_n^2) ,
 \end{equation}
 where $y_n$ is the observed data, $m_n$ is the modeled data,  and $s_n^2$ is:
\begin{equation}
    s_n^2 = \sigma_n^2 + \beta^2(m_n^2),
\end{equation}
where $\sigma_n$ are the nominal errors to the observed data.

The new Bayes factor we find is $K_{\rm{Bayes}} = 10^{1.2}$, which indicates the SN model is moderately, but not conclusively, preferred. We report the median parameters with their 1$\sigma$ errors, along with the best-fit parameters which minimize the log-likelihood function in Table \ref{modelselectiontable}. The best-fit models are shown in Figure \ref{best-fitmodels}, along with their associated $\chi$ values. According to the model selection analysis, the optimal error-stretching factor increases the error bars by $\sim 3\%$ with respect to the model value at the observed time, for both the afterglow+host and SN model. We also calculated the Bayesian evidences for the two models, while incorporating independent error stretching factors ($\beta_i$) for each telescope rather than a single factor across all data sets. We did so to investigate if the way we account for systematic uncertainties plays a role in biasing the model preferences. We still find similar results, as the SN model is favored by a Bayes factor of $K_{\rm{Bayes}} = 10^{0.7}$.

 Next, we analyze the LDT and LT photometry separately using their nominal error bars, to identify if the combination of photometry from different telescopes plays a role in biasing the model preferences. We do not fit the Gemini-South and GIT photometry separately, as there are only two Gemini-South photometry points, and the GIT photometry are all at early times where any SN excess contributing to the afterglow would be negligible. We find that the afterglow+host model is favored with a Bayes factor of $K_{\rm{Bayes}} = 10^{0.9}$ for the LDT photometry, while the two models are indistinguishable for the LT photometry. Figure \ref{afterglowphotometry} shows that the majority of the LDT photometry is at late times, where excess flux from a SN would be identifiable. This, in addition to only five and six photometry points being available respectively in the r$'$ and i$'$ bands leads to the model selection converging towards a shallow power-law decay slope in its fitting procedure, in order to fit for the excess flux. The best-fit power-law decay index is $\alpha = 1.36$ (median $\pm \, 1\sigma = 1.36 \pm 0.02$) , which is shallower than what is derived from the early-time optical photometry fitting in \S \ref{oag}. Therefore, although the afterglow+host model is preferred for the LDT data, the lack of early-time data biases the fit towards a power-law decay index that is shallower than expected  in order to fit for the excess flux. 

\begin{deluxetable*}{cccccccccc}
\tabletypesize{\footnotesize}
\tablecolumns{8}
\tablewidth{0pt}
\tablehead{\colhead{}    & \colhead{$a_{\rm{r'}}$} & \colhead{$a_{\rm{i'}}$} & \colhead{$\alpha$} & 
\colhead{$f_{\nu, \, \rm{r'}}^{\rm{host}}$} & \colhead{$f_{\nu, \, \rm{i'}}^{\rm{host}}$} & \colhead{$k_{\rm{SN\,1998bw}}$} & \colhead{$s_{\rm{SN\,1998bw}}$} & \colhead{$\rm{ln}$$(\beta)$} \\
\colhead{}    & \colhead{($\rm{\mu Jy}$)} & \colhead{$a_{\rm{i',\,AG}}$} & \colhead{} & 
\colhead{($\rm{\mu Jy}$)} & \colhead{($\rm{\mu Jy}$)} & \colhead{} & \colhead{} & \colhead{}}
 \startdata 
 Afterglow+Host (Best-fit) & $146$ & $388$ & $1.43$ &  $0.44 $  & $1.64$  & - & - & $-3.37$ \\
 Afterglow+Host (Median $\pm 1\sigma$) & $144 \pm 2$ & $384^{+5}_{-6}$ & $1.42 \pm 0.01$ &  $0.41 \pm 0.03 $  & $1.33 \pm 0.11$  & - & - & $-3.30 \pm 0.21$ \\
  Afterglow+Host+SN (Best-fit) & $147$ & $394$ & $1.46$ &  $0.35 $  & $1.17$  & $0.39$ & $0.69$ & $-3.51$ \\
  Afterglow+Host+SN (Median $\pm 1\sigma$) & $146 \pm 2 $  & $395 ^{+5}_{-6}$ & $1.47 \pm 0.01$ &  $0.39 \pm 0.03 $& $1.12^{+0.11}_{-0.12} $  & $0.43 \pm 0.10$  & $0.63 \pm 0.09$& $-3.50^{+0.23}_{-0.24}$ \\
\enddata
\caption{Bayesian model selection fitting parameters for the afterglow+host and the afterglow+host+SN model.}
\label{modelselectiontable}
\end{deluxetable*}
On the other hand, the majority of LT data is at early times and at a significantly higher cadence than the LDT photometry. The best-fit power-law decay index for the afterglow+host model is $\alpha = 1.46$ (median $\pm \, 1\sigma = 1.45 \pm 0.01$), which is consistent with the power-law decay index derived in \S \ref{oag}. The two models are indistinguishable due to a few photometry points that do show excess emission at later times, which the afterglow model cannot account for. Therefore, it is necessary to incorporate a combination of data sets in the modeling, in order to have optimal temporal coverage such that a sufficient amount of photometry points at both early and late times are accounted for in the fitting. After accounting for S-corrections modeled through our error-stretching factor, it is clear that the most accurate physical description of the LC comes from incorporating the entire data set, which we find favors the SN model moderately.

As a final sanity check, we perform a similar analysis using the observed light curve of SN 2013dx associated with GRB 130702A \citep{Delia2015, Toy2016} with the extinction values described above, again optimizing the likelihood function and allowing for an error-stretching factor. Due to the similar redshift ($z = 0.145$) to GRB 221009A, this avoids the requirement of calculating K-corrections \citep{Fulton2023,Shrestha+2023}. In this case we find a comparable Bayes factor to the analysis using SN 1998bw: $K_{\rm{Bayes}} = 10^{0.7}$, preferring the SN model. Thus we infer that the SN model preference is relatively insensitive to the details of the template SN used. We also emphasize that the detailed value of the host extinction adopted has little impact on the model selection. Since the SN flux is scaled by the free parameter $k_{\rm{SN}}$, increasing or decreasing the host extinction is largely offset in the modeling by a corresponding change in $k_{\rm{SN}}$. 

Our preference for the SN model agrees with \citet{Fulton2023}, as they find significant evidence of excess emission in the optical afterglow that is well-modeled by an additional SN component. However, they did not account for any host galaxy emission in their analysis, and Figure~\ref{afterglowphotometry} shows that the host galaxy makes non-negligible contributions to the optical afterglow at late times.  This is likely why they were able to find significant evidence of excess emission, while our preference for the SN model is moderate. \citet{Shrestha+2023} and \citet{Levan2023} report no evidence for  \textit{bright} SN emission, while \citet{Kann+2023} does not find any strong evidence for or against SN emission. None of them rule out the possibility of a faint associated SN, and one of the conclusions of \citet{Levan2023} is that an associated SN to GRB 221009A must be either substantially ($\sim 10 - 40 \%$) fainter  or bluer than SN 1998bw. Our findings point towards the former being true, as the best-fit flux-scaling factor for SN 2022xiw with respect to SN 1998bw is $k_{\rm{SN\, 1998bw}} = 0.39$, with an upper limit at the 99\% confidence interval of  $k_{\rm{SN\, 1998bw}} < 0.67$.


\section{Supernova Parameter Estimation}
\label{LCModeling}
\begin{figure*}
    \centering
\includegraphics[width=0.7\linewidth]{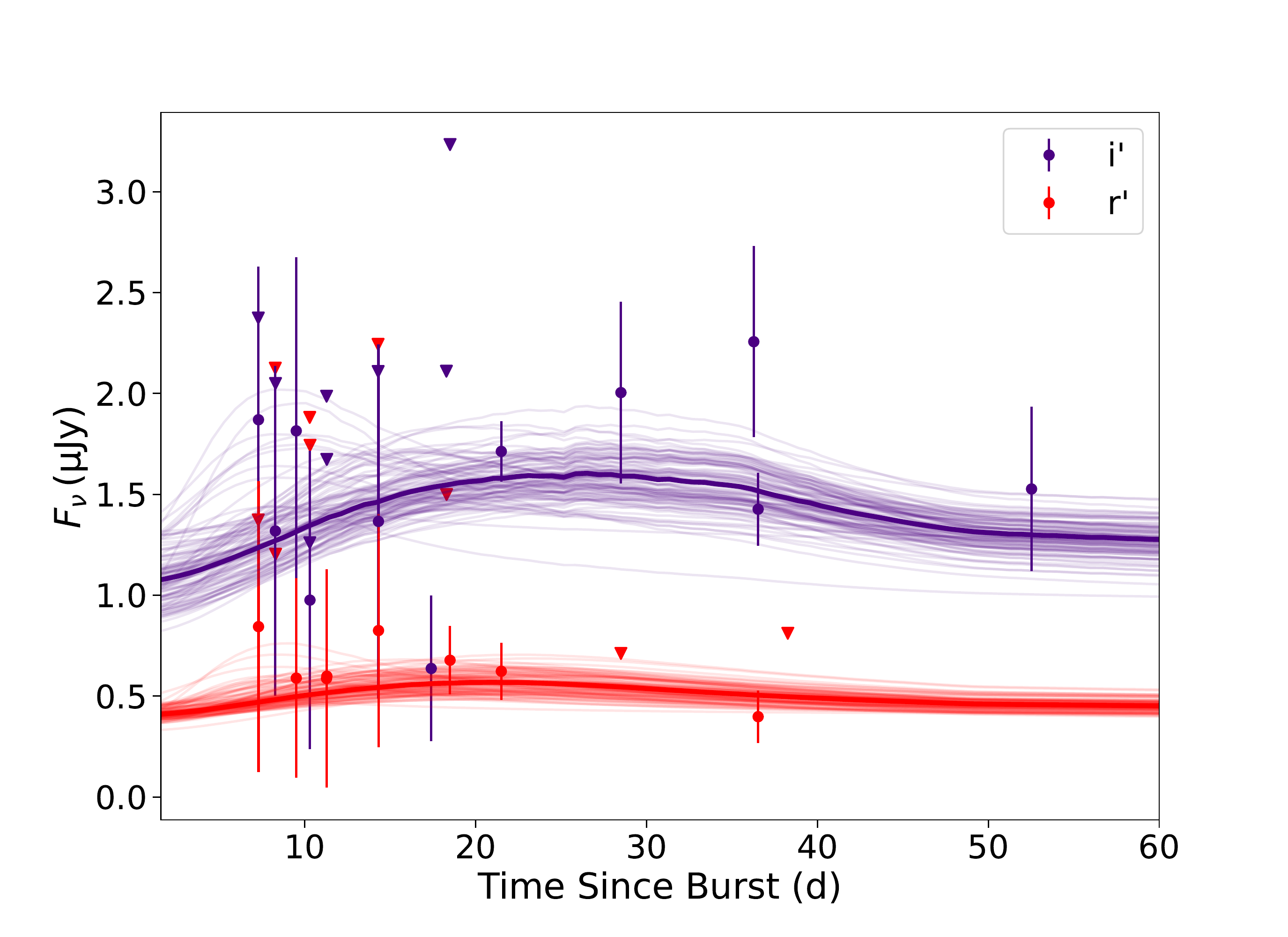}
\includegraphics[width=0.65\linewidth]{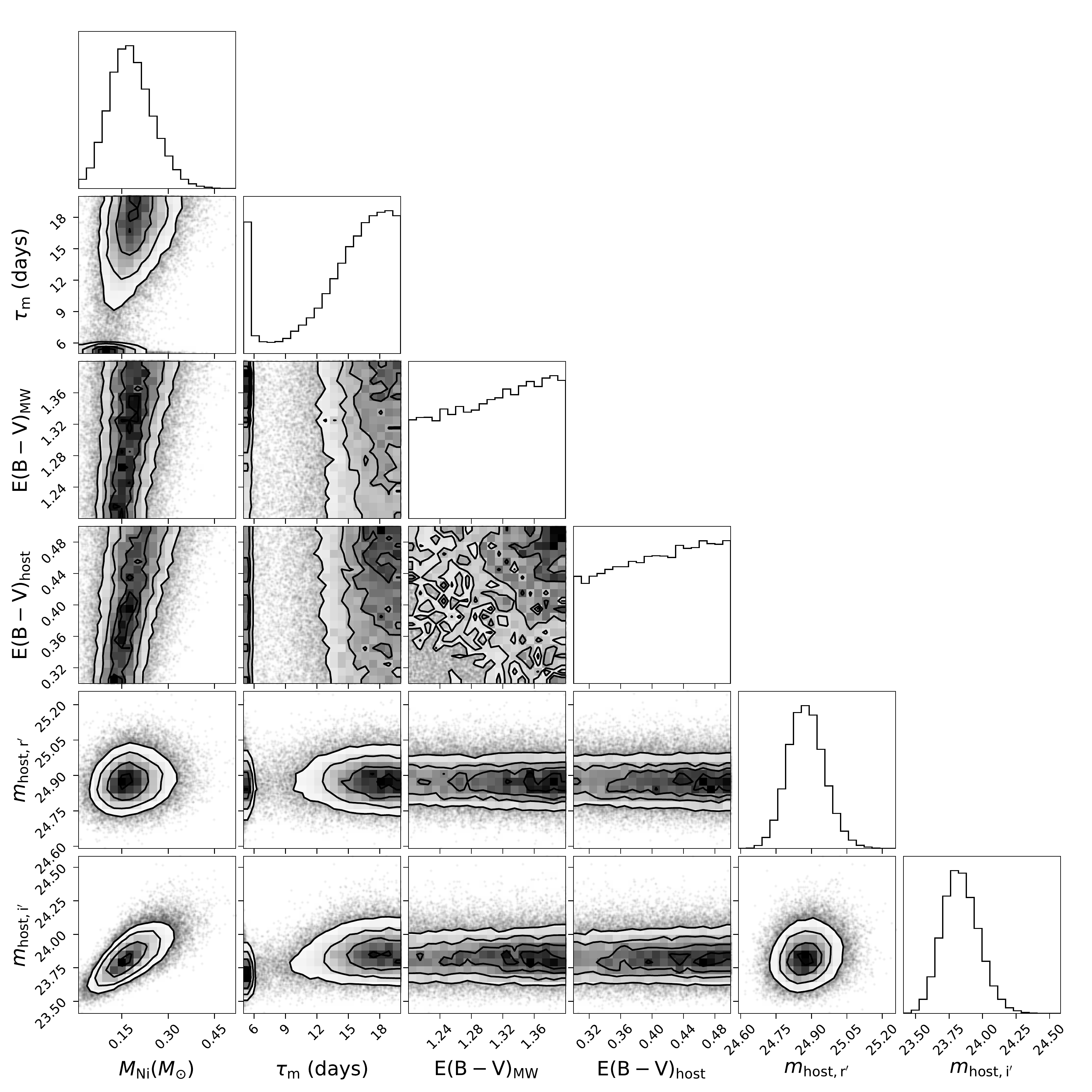}

    \caption{\textit{Top panel}: Best-fit r$'$ and i$'$-band LCs in flux space extrapolated from the best-fit bolometric LC constructed from the A82 model (details in \S \ref{LCModeling}), assuming the host galaxy extinction is a free parameter between $E(B-V)_{host} = 0.3 $ -- $ 0.5$\, mag. The best-fit values are $M_{\rm{Ni}} = 0.18^{+0.07}_{-0.06} \, \rm{M_{\odot}}$, $\tau_{\rm{m}} = 15.98^{+2.77}_{-5.36}$ days, $\rm{E(B-V)_{MW}} = 1.31 ^{+0.06}_{-0.07} $ mag, $\rm{E(B-V)_{host}} = 0.41^{+0.06}_{-0.07}$ mag, $m_{\rm{host, r'}} =  24.88 ^{+0.08} _{-0.07}  $ mag, and $m_{\rm{host, i'}} = 23.83^{+0.14}_{-0.12}$ mag. 100 random a posteriori possible models from the MCMC fitting samples are plotted,, along with the best fits in bold. \textit{Bottom panel}:  Corner plots associated with the MCMC fits of the A82 model, corresponding to the top panel. A total of 33,250 samples were generated in the posteriors.}
    \label{Mnifigure}
\end{figure*}
\subsection{Nickel Mass Estimates}
\label{sec:nickel}
After demonstrating a preference for models including a SN component, we derive flux measurements for SN 2022xiw by subtracting the the best-fit optical afterglow model (see \S \ref{oag}) from the observed r$'$ and i$'$-band photometry. We only use photometry starting from $T_0 + 7 \, \rm{d}$, and convert negative flux values after the subtractions to 3$\sigma$ upper limits. The resulting SN LC is shown in Figure~\ref{Mnifigure}. The SN photometry shown in the Figure is not host-subtracted, because we allow the host galaxy emission to vary as a free parameter within Gaussian priors corresponding to the values from \citet{Levan2023} when extracting physical parameters from the LC.
 
The decay of $^{56}\rm{Ni}$ to \Co~ and to \Fe~ releases the energy that powers the optical LC of Type I SNe, so the $^{56}\rm{Ni}$ mass is a key physical parameter that can provide insight into properties of the explosion and progenitor (\citealt{Arnett1982}; hereafter A82). Therefore, we fit semi-analytic LC models from A82 to the observed r$'$ and i$'$-band photometry after day 7 to constrain the $^{56}\rm{Ni}$ mass. Equation~36 in A82 provides an analytic expression for the bolometric luminosity of Type I SNe,  assumming full $\gamma$-ray trapping of the ejecta along with further radioactive inputs \citep{Valenti2008}. We use the infrastructure from the Hybrid Analytic Flux FittEr for Transients (HAFFET; \citealt{Yang+2023}) to perform the fits, where the two free parameters are the Nickel mass ($M_{\rm Ni}$) and the photon diffusion timescale ($\tau_{\rm{m}}$). 

Given a bolometric LC from the model, it is necessary to extract associated r$'$ and i$'$-band LCs to compare to the observed photometry. It is possible to derive light curves in individual bands from a bolometric magnitude LC using bolometric correction (BC) coefficients:
\begin{equation}
  \text{BC}_{x} = M_{bol} - M_{x},
\label{eq:bc}
\end{equation}
where $x$ is the relevant filter. 

For stripped-envelope SNe, \citet{Lyman2014} derive a g-band BC coefficient of:
\begin{equation}
  \text{BC}_{g} = 0.054 - 0.195 \times (g-r) - 0.719 \times (g-r)^{2} .
  \label{eq:bc_se_sl}
\end{equation}
 
 Here we assume that the color evolution of SN 2022xiw is identical to that of SN 1998bw. We use the BC coefficient, along with the color evolution of SN 1998bw, to generate r$'$ and i$'$-band LCs to fit to the SN photometry. First, we convert the BVRI photometry of SN 1998bw from \citet{98bwpaper} to SDSS filters using conversions from \citet{Jester2005}. At the time of each observation, we compute BC$_{g}$, along with g-r and g-i colors. Given a bolometric absolute magnitude LC from A82, we derive a g-band absolute magnitude LC from BC$_{g}$. We then compute the associated r and i-band LCs from the g-r and g-i colors derived above. Finally we apply the distance modulus, host galaxy emission, and extinction corrections described below, along with conversions from r and i band to r$'$ and i$'$ band\footnote{\url{https://classic.sdss.org/dr5/algorithms/jeg_photometric_eq_dr1.html##usno2SDSS}}, to produce observed r$'$ and i$'$-band LCs for comparison with the data. Best fit models with associated uncertainties are generated using MCMC techniques.

Given the large uncertainty in the host extinction (\S \ref{previousbroad}), we perform the fitting under three different assumptions: 1) $E(B-V)_{\mathrm{host}}$ allowed to freely vary between 0.3 and 0.5 mag; 2) $E(B-V)_{\mathrm{host}}$ fixed to a value of 0.3\,mag; and, 3) $E(B-V)_{\mathrm{host}}$ fixed to a value of 0. For all three scenarios we allow  $M_{\rm{Ni}}$, $\tau_{\rm{m}}$, and $\rm{E(B-V)_{MW}}$ to vary as additional free parameters. We allow $\rm{E(B-V)_{MW}}$ to vary due to the high uncertainty in the Galactic extinction at the location of GRB 221009A, and adopt uniform priors corresponding to the minimum and maximum values from \citet{Schlafly2011}. Additionally, we allow the host galaxy emission in the r$'$ and i$'$ bands to vary as free parameters, with Gaussian priors corresponding to the values presented in \citet{Levan2023}. We adopt uniform priors for $M_{\rm{Ni}}$ and $\tau_{\rm{m}}$ based on values from previous Type Ic-BL SNe studies \citep{Taddia2018, Corsi2016, Corsi2022}. 

 We note that there are varying interpretations of the intrinsic extinction and hydrogen column density of the host that are model-dependent on the spectral shape of the afterglow (e.g. $E(B-V)_{\rm{host}} 
< 0.1 \rm{\, mag}$; \citealt{OConnor+2023}, $N_{\rm{H}} \approx 4 \times 10^{21}\, \rm{cm^{-2}}$ ; \citealt{Tiengo2023}). However, we believe that the most likely physically plausible situation is the first where $E(B-V)_{\rm{host}}$ is allowed to vary freely between 0.3 and 0.5 mag, due to the possibility of a significant amount of intrinsic hydrogen column density in the host that may be up to $N_{\rm{H}} \approx 1.29 \times 10^{22}\, \rm{cm^{-2}}$  \citep{Williams+2023}. We present the best-fit model from this scenario and 100 random samples from the posterior distribution in Figure~\ref{Mnifigure}, along with the associated corner plots for each parameter. 

The best-fit values derived in all three scenarios are displayed in Table~\ref{physicalparameters}, and we find $M_{\rm{Ni}} = 0.05 $ -- $ 0.25 \, \rm{M_{\odot}}$, depending on the scenario. Because the SN flux is only marginally detectable with respect to the afterglow and host galaxy emission, these best-fit values should be taken with a grain of salt. However, we can more robustly determine an upper limit, and find that at the 99\% confidence level, that $M_{\rm{Ni}} < 0.36 \, \rm{M_{\odot}}$. The $M_{\rm{Ni}}$ we find is systematically lower than that of SN 1998bw ($M_{\rm{Ni}} = 0.3$--$0.9 \, \rm{M_{\odot}}$; \citealt{Sollerman2000}), which agrees with the results from \S \ref{ModelSelection}, that SN 2022xiw must be substantially fainter than SN 1998bw.

Finally, we note that if we repeat the analysis under the same assumptions as \citet{Fulton2023}, who do not explicitly account for host galaxy emission, and assume anadditional 0.8 mags of extinction in the optical when compared to the nominal galactic extinction from \citet{Schlafly2011}, we find $M_{\rm{Ni}} =  0.59 \pm 0.04 \, \rm{M_{\odot}}$. This $M_{\rm{Ni}}$ is lower, but marginally consistent with their value of $M_{\rm{Ni}} = 1.0^{+0.6}_{-0.4} \, \rm{M_{\odot}}$. The lower $M_{\rm{Ni}}$ in comparison to their work is expected, due to their use of a steeper optical afterglow power-law decay slope ($f_\nu \propto t^{-1.556 \pm 0.002}$, see \S \ref{previousbroad}), which in turn leads to more luminous SN emission, and a higher $M_{\rm{Ni}}$. 

\begin{deluxetable*}{ccccccccc}
\tabletypesize{\footnotesize}
\tablecolumns{8}
\tablewidth{0pt}
\tablehead{\colhead{} & \colhead{$M_{\rm{Ni}}$} & \colhead{$\tau_{\rm{m}}$}   & \colhead{$E(B-V)_{\rm{MW}}$} & \colhead{$E(B-V)_{\rm{host}}$} & \colhead{$M_{\rm{ej}}$} & \colhead{$E_{\rm{KE}}\, $ }             & \colhead{$m_{\rm{host, r'}}$} & \colhead{$m_{\rm{host, i'}}$} \\
\colhead{} & \colhead{$(M_{\odot})$} & \colhead{(days)}   & \colhead{(mag)} & \colhead{(mag)} & \colhead{$(M_{\odot})$} & \colhead{$(\rm{erg})$ }             & \colhead{(mag)} & \colhead{(mag)}
}

 \startdata 
        $E(B-V)_{\rm{host}} = 0.3$ -- $ 0.5 $ mag & $0.18^{+0.07}_{-0.06} $      & $15.98^{+2.77}_{-5.36}$ & $1.31 ^{+0.06}_{-0.07}$  & $0.41^{+0.06}_{-0.07}$     & $7.93^{+2.99}_{-4.43} $     & $3.71 ^{+1.40}_{-2.07} \times 10^{52} $ & $24.88 ^{+0.08}_{-0.07} $ & $23.83^{+0.14}_{-0.12}$   \\
$E(B-V)_{\rm{host}} = 0.3$ mag        & $0.14 ^{+0.05}_{-0.04} $     & $16.56^{+2.35}_{-3.24}$ & $1.31^{+0.06}_{-0.07}$   & 0.3                        & $8.53^{+2.59}_{-3.01} $     & $3.99 ^{+1.21}_{-1.41} \times 10^{52}$  & $24.88 ^{+0.08}_{-0.07} $ & $23.84^{+0.13}_{-0.12}$   \\
$E(B-V)_{\rm{host}} = 0$ mag          & $0.07 \pm 0.02 $             & $16.26^{+2.52}_{-3.34}$ & $1.31^{+0.06}_{-0.07}$   & 0                          & $8.22^{+2.74}_{-3.03} $ \,  & $3.84 ^{+1.28}_{-1.42} \times 10^{52} $ & $24.89 ^{+0.08}_{-0.07} $ & $23.82^{+0.13}_{-0.11} $  \\ 
 \enddata
 \caption{Best-fit physical parameters and their statistical 1$\sigma$ errors corresponding to the three different host extinction scenarios presented in \S \ref{sec:nickel}.}
 \label{physicalparameters}
\end{deluxetable*}

\subsection{Additional Explosion Properties}
\label{explosionproperties}
Given the photon diffusion timescale $\tau_{\rm{m}}$, it is possible to derive the total ejecta mass ($M_{\rm{ej}}$) of a SN, through equation A1 of \citet{Valenti2008}:
\begin{equation}
 \tau_\mathrm{m}^2 = \frac{2\kappa_\mathrm{opt} M_\mathrm{ej}}{\beta c v_\mathrm{sc}}\textrm{\! ,}
 \label{eq4}
\end{equation} 
where $\kappa_{\mathrm{opt}} = 0.07$\,cm$^{-2}$\,g$^{-1}$ is a constant, average opacity that is able to produce consistent results with hydrodynamical LC modeling of stripped-envelope SNe \citep{Taddia2018a},  $\beta = 13.8$ is a constant, $c$ is the speed of light, and $v_{\rm{sc}}$ is a scale velocity, which is set observationally to the photospheric velocity $v_{\rm{ph}}$, which is roughly related to the line velocity at the peak epoch.

Given that we assumed the color and spectral evolution of SN 2022xiw were identical to that of SN 1998bw, we also assume it has a comparable photospheric velocity: $v_{\rm{ph}} = {\rm 28,000}$\,km\,s$^{-1}$ \citep{Iwamoto98}. We note that SN 1998bw's photospheric velocity is high with respect to the population of GRB-SN in literature, which possesses an average of $v_{\rm{ph}} = 20,200 \, \rm{km \, s^{-1}}$, with a dispersion $\sigma = 8,500 \, \rm{km \, s^{-1}}$ \citep{cano2017}. However, a spectrum taken of the optical afterglow at $T_0 + 8 \, \rm{d}$ reported the possible existence of broad features with velocities slightly larger than SN 1998bw \citep{spectraGCN}. Therefore, our assumption of SN 2022xiw's peak photospheric velocity is valid, and may possibly under-represent the true photospheric velocity. We report the derived values in Table \ref{physicalparameters}, and find $M_{\rm{ej}} = 3.5$ -- $ 11.1 \, \rm{M_\odot}$.

Given $M_{\mathrm{ej}}$ and $v_{\rm{sc}}$, it is possible to derive the kinetic energy $E_{\rm{KE}}$ of the explosion, assuming that it is a constant density sphere undergoing homologous expansion \citep{Lyman2016}: 
\begin{equation}
 v_\mathrm{sc}^2 \equiv v_\mathrm{ph}^2 = \frac{5}{3}\frac{2E_\mathrm{K}}{M_\mathrm{ej}}\textrm{\! .}
 \label{eq:vsc}
\end{equation}
The derived values are again reported in Table \ref{physicalparameters}, and we find $E_{\rm{KE}} = (1.6$ -- $ 5.2) \times 10^{52} \, \rm{erg} $. These values are consistent with the values \citet{Fulton2023} derive, who find $M_\mathrm{ej} = 7.1^{+2.4}_{-1.7} \, \rm{M_\odot}$ and $E_{\rm{KE}} = (2.7$ -- $ 6.3) \times 10^{52} \, \rm{erg}$. We note that both $M_{\rm{ej}}$ and $E_{\rm{K}}$ largely depend on the photon diffusion timescale, and the corner plots in Figure \ref{Mnifigure} show that $\tau_{\rm{m}}$ is close to hitting the bounds of the priors. This is unsurprising given the nature of the data set, and the faintness of the SN with respect to the afterglow and host galaxy emission. Therefore, both the $M_{\rm{ej}}$ and $E_{\rm{KE}}$ we derive should also be considered as estimates, and we are unable to derive robust upper limits in this case due to the nature of the posterior of $\tau_{\rm{m}}$.

We note that throughout this analysis, systematic uncertainties likely arise from the assumptions made with the color and spectral evolution of SN 2022xiw being identical to that of SN 1998bw, as well as assuming that explosion is undergoing homologous expansion, as the presence of a relativistic jet likely impacts the spherical symmetry of the explosion. An asymmetric explosion would likely impact $\kappa_{\rm{opt}}$ and the assumptions made in Eq. \ref{eq4} from \citet{Taddia2018}. However, despite these caveats, the statistical uncertainties we find are large and likely dominate over any of these systematic uncertainties. 

\begin{figure*}
    \centering
\includegraphics[width=0.49\linewidth]{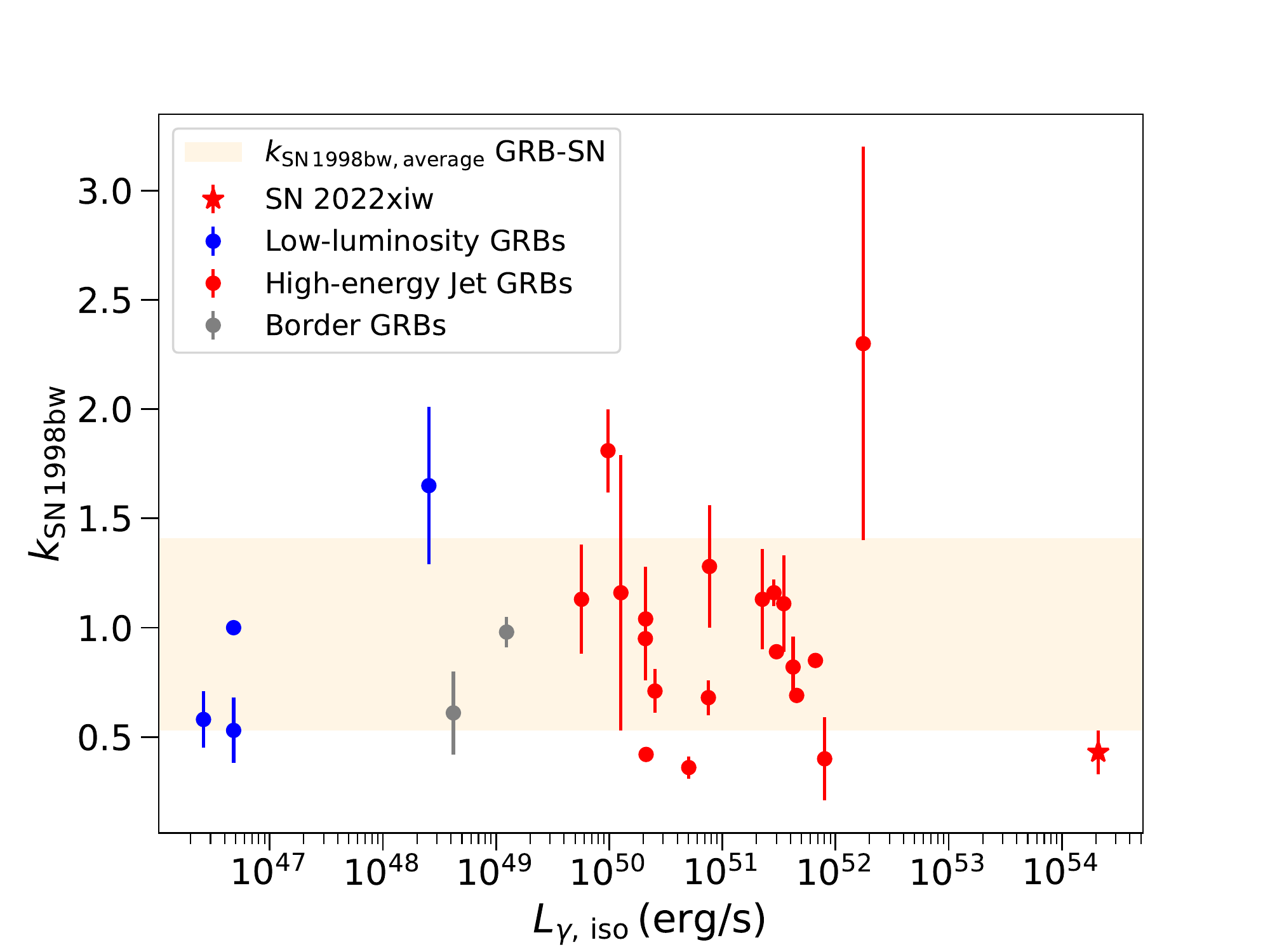}
    \includegraphics[width=0.49\linewidth]{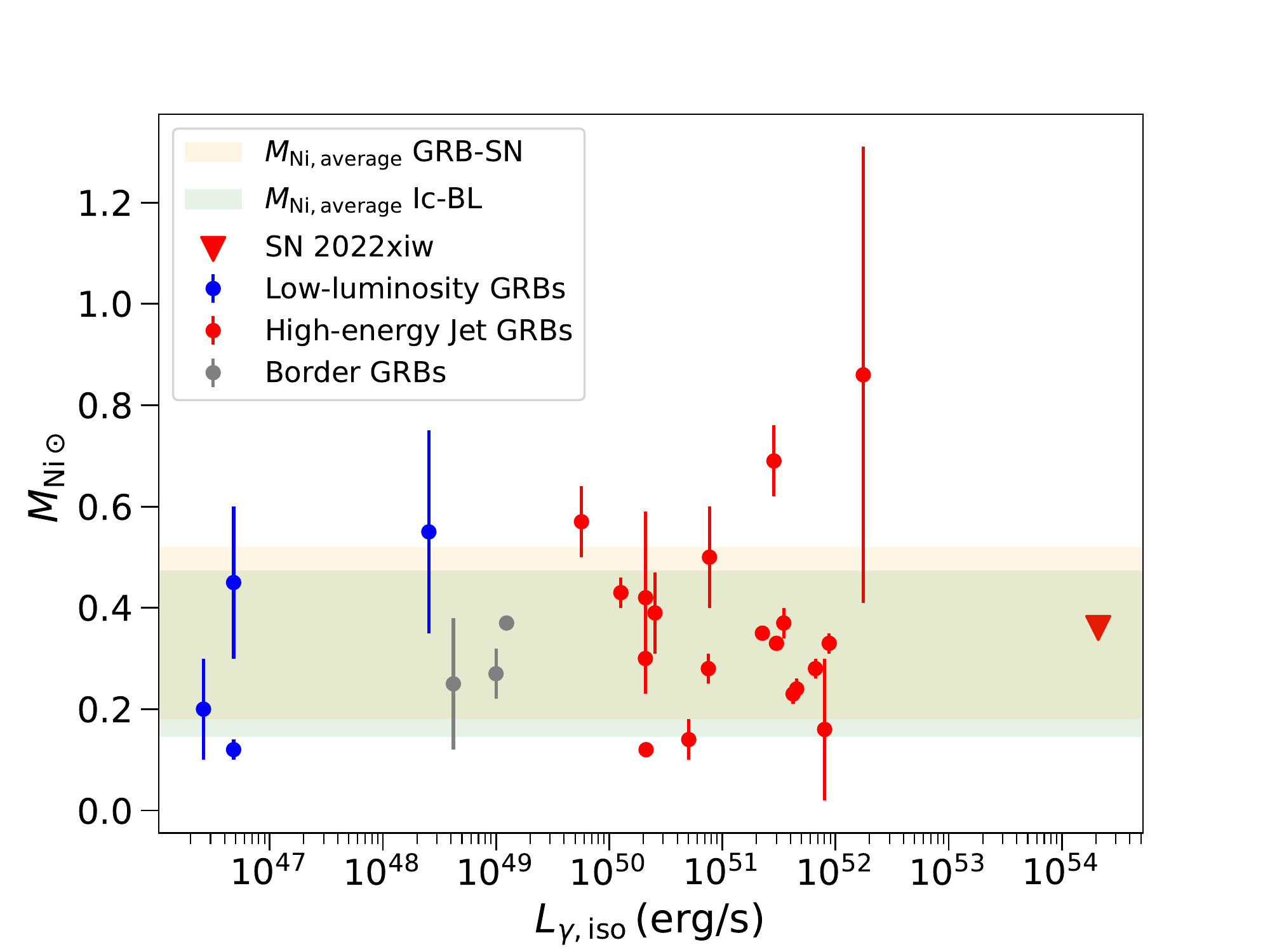}
    \caption{\textit{Left panel}: Modification of Fig. 3 from \citet{hjorth2013}, where the flux-stretching factor of SNe with respect to SN 1998bw ($k_{\rm{SN \, 1998bw}}$) is plotted against the isotropic equivalent $\gamma$-ray luminosity  ($L_{\gamma, \rm{iso}}$) for their associated GRBs. We distinguish between low-luminosity GRBs, high-energy jet GRBs, and GRBs in the border regime. We indicate the results from this work, GRB 221009A/SN 2022xiw with a star. We also show the average $k_{\rm{SN \, 1998bw}}$ for GRB-SN in the plot with the exception of GRB 2201009A/SN 2022xiw, along with its dispersion. \textit{Right panel}: A similar modification of Fig. 3 from \citet{hjorth2013}, where the $M_{\rm{Ni}}$ in $M_{\odot}$ of SNe is plotted against the isotropic equivalent $\gamma$-ray luminosity  ($L_{\gamma, \rm{iso}}$) for their associated GRBs. We make the same distinctions between GRBs as in the above panel, and label the results from GRB 221009A/SN 2022xiw with an upper limit. We also show the average $M_{\rm{Ni}}$  and dispersion for the GRB-SN plotted with the exception of GRB 221009A/SN 2022xiw, along with that of the Type Ic-BL sample not associated with GRBs from \citet{Taddia2018}.}
\label{GRBSNecomparison}
\end{figure*}

\section{Comparison to other GRB-SN}
\label{Comparison}
Based on our previous modeling, we attempt to contextualize SN 2022xiw with respect to the overall GRB-SN population.  The range of $M_{\rm{Ni}}$ found for previous GRB-SN, derived under the same assumption that the luminosity is powered by the radioactive decay of $^{56}$Ni, has an average value of $M_{\rm{Ni}} = 0.37 \, \rm{M}_{\odot}$  with a dispersion $\sigma = 0.20 \,\rm{M}_{\odot}$ \citep{cano2017}. The $M_{\rm{Ni}}$ we derive for the scenarios where we take into account host galaxy extinction ($M_{\rm{Ni}} = 0.10$--$0.25 \, M_{\odot}$) are within this range, while the $M_{\rm{Ni}}$ we derive for the scenario with zero host galaxy extinction is outside of this range.  The upper limit for the 
$M_{\rm{Ni}}$ we derive ($<0.36 \, M_{\odot}$) shows that despite GRB 221009A's highly energetic nature, its associated SN does not possess an exceptional $M_{\rm{Ni}}$  in comparison with the overall GRB-SN population- in fact, its $M_{\rm{Ni}}$  is likely lower with respect to the average. For the ejecta mass and kinetic energy, the average values inferred for previous GRB-SN are  $M_\mathrm{ej} = 6 \,  \rm{M_\odot}$ with a dispersion $\sigma =  4 \,\rm{M_\odot}$ and $E_{\rm{KE}} = 2.5 \times 10^{52} \, \rm{erg}$, with a dispersion $\sigma = 1.8 \times 10^{52} \, \rm{erg}$ \citep{cano2017}. 
Despite the caveat mentioned in \S \ref{explosionproperties} with the values we derive, both the ejecta masses and kinetic energies for all scenarios are well within these values. 

Through these comparisons, we see that SN 2022xiw possesses explosion properties that are overall broadly consistent with the GRB-SN population. This is despite its highly energetic relativistic ejecta, as it possesses an isotropic-equivalent peak $\gamma$-ray luminosity of $L_{\rm{\gamma, iso}}= 2.1\times10^{54} \, \rm{erg\, s^{-1}}$ \citep{Frederiks2023}, making it the most luminous GRB-SN ever detected \citep{cano2017, Burns2023}. Figure 3 in \citet{hjorth2013} suggests that a relationship may exist between the energy release of GRBs and their associated SN brightness. In their Figure, they plot the $L_{\rm{\gamma, iso}}$ of GRBs against the peak absolute magnitude in V band ($M_{\rm{V}}$) of their  associated SNe, and make the distinction between low luminosity GRBs ($L_{\gamma, \rm{iso}} < 10^{48.5} \rm{erg\, s^{-1}}$), and high-energy jet GRBs ($L_{\gamma, \rm{iso}} > 10^{49.5} \rm{erg\, s^{-1}}$). They report a possible direct relationship between $L_{\rm{\gamma, iso}}$ and $M_{\rm{V}}$ in low-luminosity GRBs, that turns over into a possible inverse relationship in the high-energy jet GRB region of the parameter space.

In Figure~\ref{GRBSNecomparison}, we recreate the results from \citet{hjorth2013} to test for these correlations, with a larger data set and a few modifications. We update the GRB-SN sample to include all GRB-SN with an A, B, or C classification from \citet{cano2017} and distinguish between low-luminosity GRBs in blue, high-energy jet GRBs in red, and events in the middle in gray. We plot $L_{\rm{\gamma, iso}}$ against the flux-stretching factor with respect to SN 1998bw ($k_{\rm{SN \, 1998bw}}$), along with $M_{\rm{Ni}}$. These two parameters are both proxies for  the brightness of the associated SN LC ($M_{\rm{V}}$) that \citet{hjorth2013} used. We use them both in tandem as they have the additional advantage that one is a directly observable feature from the LC ($k_{\rm{SN \, 1998bw}}$), while the other is a physical parameter derived from modeling the explosion ($M_{\rm{Ni}}$). We also add a few additional events to the sample of \citet{cano2017}, namely  GRB 200826A/AT 2020scz \citep{Ahumada+2021} to the $k_{\rm{SN \, 1998bw}}$ plot, and GRB 161219B/SN 2016jca \citep{Ashall+2019} and GRB 171010A/SN 2017htp \citep{Melandri2019} to the $M_{\rm{Ni}}$ plot. Finally, we add the results from this work to both plots, with the best-fit flux-stretching factor from Table \ref{modelselectiontable}, $k_{\rm{SN \, 1998bw}} = 0.39$ denoted as a star, and the limit of $M_{\rm{Ni}} < 0.36 \, M_{\odot}$ denoted as an upper limit. 

In addition, we also plot the average $k_{\rm{SN \, 1998bw}}$ ($k_{\rm{SN \, 1998bw, \, average}} = 0.97$) and dispersion ($\sigma = 0.44$) for the GRB-SN plotted with the exception of SN 2022xiw shaded in orange, and the average $M_{\rm{Ni}}$ ($M_{\rm{Ni, \, average}} = 0.35 \, \rm{M_{\odot}}$) and dispersion  ($\sigma = 0.17 \,  \rm{M_{\odot}}$) for the GRB-SN plotted with the exception of SN 2022xiw shaded again in orange, along with for the Type Ic-BL population not associated with GRBs ($M_{\rm{Ni, \, average}} = 0.31 \, \rm{M_{\odot}}, \, \sigma =  0.16 \, \rm{M_{\odot}}$;  \citealt{Taddia2018}) shaded in green. From the Figure, we notice that SN 2022xiw's $k_{\rm{SN \, 1998bw}}$ is slightly lower than the average range for the rest of the sample, despite the burst being orders of magnitude more energetic. Furthermore, SN 2022xiw's $M_{\rm{Ni}}$ is also likely on the lower end with respect to the overall population. The overlap between the average $M_{\rm{Ni}}$  in normal Type Ic-BL SNe and GRB-SNe suggest that there are no intrinsic differences between the brightness of SN associated with GRBs with those that are not.

We use the Pearson Correlation Coefficient Test to determine if there are correlations between $L_{\rm{\gamma, iso}}$ and $k_{\rm{SN \, 1998bw}}$ and $L_{\rm{\gamma, iso}}$ and $M_{\rm{Ni}}$, between the low-luminosity GRBs, high-energy jet GRBs, and the entire data set. In the low-luminosity regime, we do find a correlation for both $k_{\rm{SN \, 1998bw}}$ and $M_{\rm{Ni}}$, with coefficients of 0.90 and 0.68 respectively. However, this correlation must be taken with caution, as there are only 4 low-luminosity GRBs in the sample. When testing the high-energy jet GRBs, we find no significant evidence for any correlations in the data, with coefficients of $-0.27$ and 0.00 for $k_{\rm{SN \, 1998bw}}$ and $M_{\rm{Ni}}$ respectively. When testing the overall data set including the border GRBs, we also find no significant evidence for any correlations, with coefficients of $-0.22$ and 0.01 for $k_{\rm{SN \, 1998bw}}$ and $M_{\rm{Ni}}$ respectively. These tests done on the high-energy jet GRBs and entire data set all had p-values greater than 0.25. The lack of correlations are quite interesting, especially because $L_{\rm{\gamma, iso}}$ ranges over 7 orders of magnitude. This suggests that SN emission appears to be largely decoupled from any central engine activity in GRB-SN systems.


\section{Conclusion}
\label{Conclusion}
By modeling the optical emission from GRB 221009A, we find moderate, but not conclusive statistical evidence ($K_{\rm{Bayes}} = 10^{1.2}$) for the presence of  associated supernova emission, and find that GRB 221009A's associated SN 2022xiw must be substantially fainter than SN 1998bw. We also extract physical parameters associated with the SN, assuming three different host galaxy extinction scenarios: 1) $E(B-V)_{host}$ is allowed to vary as a free parameter between 0.3 and 0.5 mag;  2) $E(B-V)_{host}$ = 0.3 mag; and 3) there is no host galaxy extinction. The most physically plausible scenario is the first, as there is evidence for extinction larger than the nominal \citet{Schlafly2011} value, though with a large degree of associated uncertainty \citep{Williams+2023,Kann+2023,Levan2023}.

We derive $M_{\rm{Ni}} = 0.05$ -- $0.25 \, \rm{M_\odot}$, $M_{\rm{ej}} = 7.2$ --$11.8 \, \rm{M_\odot}$, and $E_{\rm{KE}} = (3.3$ -- $ 5.5) \times 10^{52} \, \rm{erg}$. These values are weakly constrained due to the faintness of the SN emission with respect to the afterglow and host emission, but we robustly constrain an upper-limit on the $M_{\rm{Ni}}$ of $M_{\rm{Ni}} < 0.36 \,  \rm{M_\odot}$. All of the explosion parameters lie within the range of those found in previous GRB-SN in literature \citep{cano2017}, suggesting that even the most extreme GRBs can produce SNe with explosion properties typical of the overall GRB-SN population. We investigate the explosion parameters with respect to the overall GRB-SN population, and find that there is no significant correlation between the luminosity of GRBs and their associated SN's brightness. This suggests that central engine activity in GRB-SN systems and SN emission are largely decoupled, and further studies in the future pinpointing why will be of utmost importance to unraveling the GRB-SN connection in totality. This event is an important addition to the GRB-SN population, and is a  prime example for why it is extremely important to continue analyzing high-energy GRBs to further understand the GRB-SN connection.

\section*{Acknowledgements} 
G.~P.~S.~acknowledges Geoffrey Ryan for useful discussions about Bayesian modeling techniques, as well as Mansi Kasliwal, Kishalay De, Ryan Lau, Viraj Karambelkar, Michael Ashley, Jacob Jencson, and the Palomar Gattini Infrared Survey team for discussions about the paper during the drafting process. G.~P.~S. thanks
James Bauer and Quanzhi Ye for their assistance obtaining LDT observations, and Simi Bhullar for her moral support throughout the paper writing process. M.~W.~C. acknowledges support from the National Science
Foundation with grant numbers PHY-2010970 and OAC-2117997.
R.~S. acknowledges support from grant number 12073029 from the National Natural Science Foundation of China (NSFC).
J.~H.~G. and E.~T. acknowledge support from the European Research Council (ERC) under the European Union’s Horizon 2020 research and innovation programme, grant 101002761. A. J. D. was supported, and M. C. M. was supported in part, by NASA ADAP grant 80NSSC21K0649 The material is based upon work supported by NASA under award number 80GSFC21M0002. 

These results made use of Lowell Observatory’s Lowell Discovery Telescope (LDT),
formerly the Discovery Channel Telescope. Lowell operates the LDT in partnership with
Boston University, Northern Arizona University, the University of Maryland, and the University of Toledo. Partial support of the LDT was provided by Discovery Communications. LMI was built by Lowell Observatory using funds from the National Science Foundation (AST-1005313). These results were also based on observations obtained at the international Gemini Observatory, a program of NSF’s NOIRLab, which is managed by the Association of Universities for Research in Astronomy (AURA) under a cooperative agreement with the National Science Foundation on behalf of the Gemini Observatory partnership: the National Science Foundation (United States), National Research Council (Canada), Agencia Nacional de Investigaci\'{o}n y Desarrollo (Chile), Ministerio de Ciencia, Tecnolog\'{i}a e Innovaci\'{o}n (Argentina), Minist\'{e}rio da Ci\^{e}ncia, Tecnologia, Inova\c{c}\~{o}es e Comunica\c{c}\~{o}es (Brazil), and Korea Astronomy and Space Science Institute (Republic of Korea). The GROWTH India Telescope (GIT) is a 70-cm telescope with a 0.7-degree field of view, set up by the Indian Institute of Astrophysics (IIA) and the Indian Institute of Technology Bombay (IITB) with funding from Indo-US Science and Technology Forum and the Science and Engineering Research Board, Department of Science and Technology (DST), Government of India.  It is located at the Indian Astronomical Observatory (Hanle), operated by IIA an autonomous institute under DST. We acknowledge funding by the IITB alumni batch of 1994, which partially supports operations of the telescope.


\vspace{5mm}
\facilities{Lowell Discovery Telescope (LDT), GROWTH-India Telesceope (GIT), Gemini-South Telescope}

\software{{\sc PyMultinest} \citep{Feroz09, Buchner+2014};
          {\sc astropy} \citep{astropy};
          {\sc emcee} \citep{emcee}.
          }
\clearpage
\appendix
\begin{longtable}[c]{cccc}
\hline
\hline
$t_{\rm{obs}} - T_0$ (days) & Filter & AB mag  & Uncertainty \\ 
\hline
        0.128349 & g & 17.66 & 0.07 \\ 
        0.211 & g & 18.22 & 0.33 \\ 
        1.134 & g & 20.43 & 0.2 \\ 
        1.152193 & g & 20.13 & 0.08 \\ 
        1.165 & g & 20.53 & 0.11 \\ 
        2.166539 & g & 21.15 & 0.21 \\ 
        1.02802141 & g & 20.037 & 0.205 \\ 
        1.03149415 & g & 20.248 & 0.234 \\ 
        1.12791174 & g & 20.228 & 0.339 \\ 
        1.13782442 & g & 20.299 & 0.363 \\ 
        1.1412292 & g & 20.634 & 0.469 \\ 
        1.27027935 & g & 20.33 & 0.06 \\ 
        1.39404935 & g & 20.41 & 0.12 \\ 
        2.29360935 & g & 21.05 & 0.12 \\ 
        3.28066935 & g & 21.61 & 0.19 \\ 
        6.09 & g & 22.61 & 0.12 \\ 
        11.55024 & g & 23.7 & 0.2 \\ 
        0.131049 & r & 16.16 & 0.07 \\ 
        0.211 & r & 16.76 & 0.08 \\ 
        0.43625 & r & 17.36 & 0.12 \\ 
        1.136 & r & 18.57 & 0.05 \\ 
        1.159669 & r & 18.65 & 0.02 \\ 
        1.168 & r & 18.64 & 0.03 \\ 
        1.172 & r & 18.43 & 0.11 \\ 
        1.2568 & r & 18.74 & 0.12 \\ 
        1.301 & r & 18.96 & 0.1 \\ 
        1.314699 & r & 18.81 & 0.05 \\ 
        2.138926 & r & 19.53 & 0.04 \\ 
        2.306539 & r & 19.67 & 0.11 \\ 
        3.156539 & r & 20.03 & 0.06 \\ 
        3.176539 & r & 19.97 & 0.08 \\ 
        3.206539 & r & 20.07 & 0.19 \\ 
        3.226539 & r & 20.32 & 0.17 \\ 
        3.2459 & r & 20.23 & 0.09 \\ 
        3.266539 & r & 20.17 & 0.12 \\ 
        3.296539 & r & 20.26 & 0.16 \\ 
        3.316539 & r & 20.24 & 0.19 \\ 
        4.146539 & r & 20.53 & 0.09 \\ 
        4.176539 & r & 20.63 & 0.09 \\ 
        4.196539 & r & 20.71 & 0.15 \\ 
        4.216539 & r & 20.54 & 0.1 \\ 
        4.236539 & r & 20.55 & 0.12 \\ 
        4.266539 & r & 20.74 & 0.16 \\ 
        4.286539 & r & 20.9 & 0.23 \\ 
        4.306539 & r & 20.86 & 0.27 \\ 
        4.674099 & r & 20.92 & 0.05 \\ 
        5.7 & r & 21.13 & 0.06 \\ 
        0.214 & i & 15.58 & 0.03 \\ 
        1.154 & i & 17.56 & 0.05 \\ 
        1.167041 & i & 17.52 & 0.01 \\ 
        1.17 & i & 17.58 & 0.01 \\ 
        1.322576 & i & 17.69 & 0.02 \\ 
        1.48 & i & 17.92 & 0.11 \\ 
        2.14674 & i & 18.4 & 0.02 \\ 
        2.316539 & i & 18.49 & 0.04 \\ 
        3.146539 & i & 18.82 & 0.03 \\ 
        3.176539 & i & 19.02 & 0.07 \\ 
        3.196539 & i & 19.09 & 0.1 \\ 
        3.216539 & i & 18.95 & 0.07 \\ 
        3.236539 & i & 18.93 & 0.04 \\ 
        3.2459 & i & 18.91 & 0.11 \\ 
        3.266539 & i & 18.93 & 0.04 \\ 
        3.286539 & i & 18.92 & 0.04 \\ 
        4.156539 & i & 19.51 & 0.06 \\ 
        4.176539 & i & 19.41 & 0.05 \\ 
        4.206539 & i & 19.52 & 0.05 \\ 
        4.226539 & i & 19.44 & 0.04 \\ 
        4.246539 & i & 19.45 & 0.05 \\ 
        4.266539 & i & 19.43 & 0.05 \\ 
        4.296539 & i & 19.48 & 0.06 \\ 
        4.316539 & i & 19.5 & 0.07 \\ 
        4.45 & i & 19.89 & 0.05 \\ 
        4.671579 & i & 19.88 & 0.02 \\ 
        5.7 & i & 20.01 & 0.05 \\ 
        6.07 & i & 20.01 & 0.04 \\ 
        0.216667 & z & 14.89 & 0.03 \\ 
        1.166 & z & 16.93 & 0.05 \\ 
        1.171 & z & 16.87 & 0.05 \\ 
        1.174389 & z & 16.81 & 0.01 \\ 
        1.330626 & z & 16.99 & 0.01 \\ 
        1.48 & z & 16.92 & 0.11 \\ 
        2.154231 & z & 17.69 & 0.02 \\ 
        2.326539 & z & 17.72 & 0.03 \\ 
        3.166539 & z & 18.2 & 0.04 \\ 
        3.186539 & z & 18.19 & 0.05 \\ 
        3.206539 & z & 18.4 & 0.08 \\ 
        3.236539 & z & 18.26 & 0.03 \\ 
        3.2459 & z & 18.35 & 0.13 \\ 
        3.256539 & z & 18.23 & 0.03 \\ 
        3.276539 & z & 18.23 & 0.04 \\ 
        3.306539 & z & 18.3 & 0.04 \\ 
        3.326539 & z & 18.18 & 0.04 \\ 
        4.166539 & z & 18.63 & 0.05 \\ 
        4.186539 & z & 18.76 & 0.05 \\ 
        4.191 & z & 18.8 & 0.1 \\ 
        4.206539 & z & 18.69 & 0.04 \\ 
        4.236539 & z & 18.75 & 0.04 \\ 
        4.256539 & z & 18.74 & 0.05 \\ 
        4.276539 & z & 18.83 & 0.05 \\ 
        4.306539 & z & 18.74 & 0.05 \\ 
        4.326539 & z & 18.71 & 0.06 \\ 
        4.670249 & z & 19.21 & 0.02 \\ 
        5.7 & z & 19.39 & 0.05 \\ 
\hline
\caption{Optical photometry and 1$\sigma$ errors of GRB 221009A, including contributions from its afterglow, host galaxy, and associated SN 2022xiw, from publicly available GCNs. All times are in the observer frame. The magnitudes are not corrected for Galactic extinction.}
\label{GCNphot}
\end{longtable}

\bibliography{main}{}

\begin{thebibliography}{}
\expandafter\ifx\csname natexlab\endcsname\relax\def\natexlab#1{#1}\fi
\providecommand{\url}[1]{\href{#1}{#1}}
\providecommand{\dodoi}[1]{doi:~\href{http://doi.org/#1}{\nolinkurl{#1}}}
\providecommand{\doeprint}[1]{\href{http://ascl.net/#1}{\nolinkurl{http://ascl.net/#1}}}
\providecommand{\doarXiv}[1]{\href{https://arxiv.org/abs/#1}{\nolinkurl{https://arxiv.org/abs/#1}}}

\bibitem[{{Ahumada} {et~al.}(2021){Ahumada}, {Singer}, {Anand}, {Coughlin},
  {Kasliwal}, {Ryan}, {Andreoni}, {Cenko}, {Fremling}, {Kumar}, {Pang},
  {Burns}, {Cunningham}, {Dichiara}, {Dietrich}, {Svinkin}, {Almualla},
  {Castro-Tirado}, {De}, {Dunwoody}, {Gatkine}, {Hammerstein}, {Iyyani},
  {Mangan}, {Perley}, {Purkayastha}, {Bellm}, {Bhalerao}, {Bolin}, {Bulla},
  {Cannella}, {Chandra}, {Duev}, {Frederiks}, {Gal-Yam}, {Graham}, {Ho},
  {Hurley}, {Karambelkar}, {Kool}, {Kulkarni}, {Mahabal}, {Masci}, {McBreen},
  {Pandey}, {Reusch}, {Ridnaia}, {Rosnet}, {Rusholme}, {Carracedo}, {Smith},
  {Soumagnac}, {Stein}, {Troja}, {Tsvetkova}, {Walters}, \&
  {Valeev}}]{Ahumada+2021}
{Ahumada}, T., {Singer}, L.~P., {Anand}, S., {et~al.} 2021, Nature Astronomy,
  5, 917, \dodoi{10.1038/s41550-021-01428-7}

\bibitem[{{Arnett}(1982)}]{Arnett1982}
{Arnett}, W.~D. 1982, \apj, 253, 785, \dodoi{10.1086/159681}

\bibitem[{{Ashall} {et~al.}(2019){Ashall}, {Mazzali}, {Pian}, {Woosley},
  {Palazzi}, {Prentice}, {Kobayashi}, {Holmbo}, {Levan}, {Perley},
  {Stritzinger}, {Bufano}, {Filippenko}, {Melandri}, {Oates}, {Rossi},
  {Selsing}, {Zheng}, {Castro-Tirado}, {Chincarini}, {D'Avanzo}, {De Pasquale},
  {Emery}, {Fruchter}, {Hurley}, {Moller}, {Nomoto}, {Tanaka}, \&
  {Valeev}}]{Ashall+2019}
{Ashall}, C., {Mazzali}, P.~A., {Pian}, E., {et~al.} 2019, \mnras, 487, 5824,
  \dodoi{10.1093/mnras/stz1588}

\bibitem[{{Astropy Collaboration} {et~al.}(2013){Astropy Collaboration},
  {Robitaille}, {Tollerud}, {Greenfield}, {Droettboom}, {Bray}, {Aldcroft},
  {Davis}, {Ginsburg}, {Price-Whelan}, {Kerzendorf}, {Conley}, {Crighton},
  {Barbary}, {Muna}, {Ferguson}, {Grollier}, {Parikh}, {Nair}, {Unther},
  {Deil}, {Woillez}, {Conseil}, {Kramer}, {Turner}, {Singer}, {Fox}, {Weaver},
  {Zabalza}, {Edwards}, {Azalee Bostroem}, {Burke}, {Casey}, {Crawford},
  {Dencheva}, {Ely}, {Jenness}, {Labrie}, {Lim}, {Pierfederici}, {Pontzen},
  {Ptak}, {Refsdal}, {Servillat}, \& {Streicher}}]{astropy}
{Astropy Collaboration}, {Robitaille}, T.~P., {Tollerud}, E.~J., {et~al.} 2013,
  \aap, 558, A33, \dodoi{10.1051/0004-6361/201322068}

\bibitem[{{Barbary} {et~al.}(2016){Barbary}, {Barclay}, {Biswas}, {Craig},
  {Feindt}, {Friesen}, {Goldstein}, {Jha}, {Rodney}, {Sofiatti}, {Thomas}, \&
  {Wood-Vasey}}]{SNCosmo}
{Barbary}, K., {Barclay}, T., {Biswas}, R., {et~al.} 2016, {SNCosmo: Python
  library for supernova cosmology}, Astrophysics Source Code Library, record
  ascl:1611.017.
\newblock \doeprint{1611.017}

\bibitem[{{Barthelmy} {et~al.}(2005){Barthelmy}, {Barbier}, {Cummings},
  {Fenimore}, {Gehrels}, {Hullinger}, {Krimm}, {Markwardt}, {Palmer},
  {Parsons}, {Sato}, {Suzuki}, {Takahashi}, {Tashiro}, \&
  {Tueller}}]{Barthelmy2005}
{Barthelmy}, S.~D., {Barbier}, L.~M., {Cummings}, J.~R., {et~al.} 2005, \ssr,
  120, 143, \dodoi{10.1007/s11214-005-5096-3}

\bibitem[{{Belkin} {et~al.}(2022{\natexlab{a}}){Belkin}, {Kim}, {Pozanenko},
  {Krugov}, {Aimuratov}, {Pankov}, \& {GRB IKI FuN}}]{GCN.32769}
{Belkin}, S., {Kim}, V., {Pozanenko}, A., {et~al.} 2022{\natexlab{a}}, GRB
  Coordinates Network, 32769, 1

\bibitem[{{Belkin} {et~al.}(2022{\natexlab{b}}){Belkin}, {Moskvitin}, {Kim},
  {Pozanenko}, {Krugov}, {Uklein}, {Pankov}, \& {GRB IKI FuN}}]{GCN.32818}
{Belkin}, S., {Moskvitin}, A., {Kim}, V., {et~al.} 2022{\natexlab{b}}, GRB
  Coordinates Network, 32818, 1

\bibitem[{{Belkin} {et~al.}(2022{\natexlab{c}}){Belkin}, {Nazarov},
  {Pozanenko}, {Pankov}, \& {IKI GRB FuN}}]{GCN.32684}
{Belkin}, S., {Nazarov}, S., {Pozanenko}, A., {Pankov}, N., \& {IKI GRB FuN}.
  2022{\natexlab{c}}, GRB Coordinates Network, 32684, 1

\bibitem[{{Bertin}(2010)}]{2010ascl.soft10068B}
{Bertin}, E. 2010, {SWarp: Resampling and Co-adding FITS Images Together}.
\newblock \doeprint{1010.068}

\bibitem[{{Bertin} \& {Arnouts}(1996)}]{1996AAS..117..393B}
{Bertin}, E., \& {Arnouts}, S. 1996, \aaps, 117, 393,
  \dodoi{10.1051/aas:1996164}

\bibitem[{{Bikmaev} {et~al.}(2022{\natexlab{a}}){Bikmaev}, {Khamitov},
  {Irtuganov}, {Gorbachev}, {Sakhibullin}, \& {Burenin}}]{GCN.32743}
{Bikmaev}, I., {Khamitov}, I., {Irtuganov}, E., {et~al.} 2022{\natexlab{a}},
  GRB Coordinates Network, 32743, 1

\bibitem[{{Bikmaev} {et~al.}(2022{\natexlab{b}}){Bikmaev}, {Khamitov},
  {Irtuganov}, {Gorbachev}, {Sakhibullin}, \& {Burenin}}]{GCN.32752}
---. 2022{\natexlab{b}}, GRB Coordinates Network, 32752, 1

\bibitem[{{Brivio} {et~al.}(2022){Brivio}, {Ferro}, {D'Avanzo}, {Fugazza},
  {Melandri}, {Covino}, \& {REM Team}}]{GCN.32652}
{Brivio}, R., {Ferro}, M., {D'Avanzo}, P., {et~al.} 2022, GRB Coordinates
  Network, 32652, 1

\bibitem[{{Buchner} {et~al.}(2014){Buchner}, {Georgakakis}, {Nandra}, {Hsu},
  {Rangel}, {Brightman}, {Merloni}, {Salvato}, {Donley}, \&
  {Kocevski}}]{Buchner+2014}
{Buchner}, J., {Georgakakis}, A., {Nandra}, K., {et~al.} 2014, \aap, 564, A125,
  \dodoi{10.1051/0004-6361/201322971}

\bibitem[{{Burns} {et~al.}(2023){Burns}, {Svinkin}, {Fenimore}, {Ag{\"u}{\'\i}
  Fern{\'a}ndez}, {Frederiks}, {Kann}, {Hamburg}, {Lesage}, {Temiraev},
  {Tsvetkova}, {Bissaldi}, {Briggs}, {Fletcher}, {Goldstein}, {Hui}, {Hristov},
  {Kocevski}, {Lysenko}, {Mailyan}, {Racusin}, {Ridnaia}, {Roberts}, {Ulanov},
  {Veres}, {Wilson-Hodge}, \& {Wood}}]{Burns2023}
{Burns}, E., {Svinkin}, D., {Fenimore}, E., {et~al.} 2023, arXiv e-prints,
  arXiv:2302.14037, \dodoi{10.48550/arXiv.2302.14037}

\bibitem[{{Cano} {et~al.}(2017){Cano}, {Wang}, {Dai}, \& {Wu}}]{cano2017}
{Cano}, Z., {Wang}, S.-Q., {Dai}, Z.-G., \& {Wu}, X.-F. 2017, Advances in
  Astronomy, 2017, 8929054, \dodoi{10.1155/2017/8929054}

\bibitem[{{Cardelli} {et~al.}(1989){Cardelli}, {Clayton}, \&
  {Mathis}}]{Cardelli1989}
{Cardelli}, J.~A., {Clayton}, G.~C., \& {Mathis}, J.~S. 1989, \apj, 345, 245,
  \dodoi{10.1086/167900}

\bibitem[{{Castro-Tirado} {et~al.}(2022){Castro-Tirado}, {Sanchez-Ramirez},
  {Hu}, {Caballero-Garcia}, {Castro Tirado}, {Fernandez-Garcia},
  {Perez-Garcia}, {Lombardi}, {Pandey}, {Yang}, \& {Zhang}}]{GCN.32686}
{Castro-Tirado}, A.~J., {Sanchez-Ramirez}, R., {Hu}, Y.~D., {et~al.} 2022, GRB
  Coordinates Network, 32686, 1

\bibitem[{{Clocchiatti} {et~al.}(2011){Clocchiatti}, {Suntzeff}, {Covarrubias},
  \& {Candia}}]{98bwpaper}
{Clocchiatti}, A., {Suntzeff}, N.~B., {Covarrubias}, R., \& {Candia}, P. 2011,
  \aj, 141, 163, \dodoi{10.1088/0004-6256/141/5/163}

\bibitem[{{Corsi} {et~al.}(2016){Corsi}, {Gal-Yam}, {Kulkarni}, {Frail},
  {Mazzali}, {Cenko}, {Kasliwal}, {Cao}, {Horesh}, {Palliyaguru}, {Perley},
  {Laher}, {Taddia}, {Leloudas}, {Maguire}, {Nugent}, {Sollerman}, \&
  {Sullivan}}]{Corsi2016}
{Corsi}, A., {Gal-Yam}, A., {Kulkarni}, S.~R., {et~al.} 2016, \apj, 830, 42,
  \dodoi{10.3847/0004-637X/830/1/42}

\bibitem[{{Corsi} {et~al.}(2022){Corsi}, {Ho}, {Cenko}, {Kulkarni}, {Anand},
  {Yang}, {Sollerman}, {Srinivasaragavan}, {Omand}, {Balasubramanian}, {Frail},
  {Fremling}, {Perley}, {Yao}, {Dahiwale}, {De}, {Dugas}, {Hankins}, {Jencson},
  {Kasliwal}, {Tzanidakis}, {Bellm}, {Laher}, {Masci}, {Purdum}, \&
  {Regnault}}]{Corsi2022}
{Corsi}, A., {Ho}, A. Y.~Q., {Cenko}, S.~B., {et~al.} 2022, arXiv e-prints,
  arXiv:2210.09536, \dodoi{10.48550/arXiv.2210.09536}

\bibitem[{{D'Avanzo} {et~al.}(2022){D'Avanzo}, {Ferro}, {Brivio}, {Bernardini},
  {Fugazza}, {Campana}, {Covino}, {D'Elia}, {De Pasquale}, {Malesani},
  {Melandri}, {Palazzi}, {Piranomonte}, {Rossi}, {Sbarufatti}, {Tagliaferri},
  {REM Team}, \& {CIBO Collaboration}}]{GCN.32755}
{D'Avanzo}, P., {Ferro}, M., {Brivio}, R., {et~al.} 2022, GRB Coordinates
  Network, 32755, 1

\bibitem[{{de Ugarte Postigo} {et~al.}(2022{\natexlab{a}}){de Ugarte Postigo},
  {Izzo}, {Thoene}, {Fynbo}, {Kann}, {Agui Fernandez}, \&
  {Tanvir}}]{spectraGCN}
{de Ugarte Postigo}, A., {Izzo}, L., {Thoene}, C.~C., {et~al.}
  2022{\natexlab{a}}, GRB Coordinates Network, 32800, 1

\bibitem[{{de Ugarte Postigo} {et~al.}(2022{\natexlab{b}}){de Ugarte Postigo},
  {Izzo}, {Pugliese}, {Xu}, {Schneider}, {Fynbo}, {Tanvir}, {Malesani},
  {Saccardi}, {Kann}, {Wiersema}, {Gompertz}, {Thoene}, {Levan}, \& {Stargate
  Collaboration}}]{XshooterGCN}
{de Ugarte Postigo}, A., {Izzo}, L., {Pugliese}, G., {et~al.}
  2022{\natexlab{b}}, GRB Coordinates Network, 32648, 1

\bibitem[{{de Wet} {et~al.}(2022){de Wet}, {Groot}, \& {Meerlicht
  Consortium}}]{GCN.32646}
{de Wet}, S., {Groot}, P.~J., \& {Meerlicht Consortium}. 2022, GRB Coordinates
  Network, 32646, 1

\bibitem[{{D'Elia} {et~al.}(2015){D'Elia}, {Pian}, {Melandri}, {D'Avanzo},
  {Della Valle}, {Mazzali}, {Piranomonte}, {Tagliaferri}, {Antonelli},
  {Bufano}, {Covino}, {Fugazza}, {Malesani}, {M{\o}ller}, \&
  {Palazzi}}]{Delia2015}
{D'Elia}, V., {Pian}, E., {Melandri}, A., {et~al.} 2015, \aap, 577, A116,
  \dodoi{10.1051/0004-6361/201425381}

\bibitem[{{Dichiara} {et~al.}(2022){Dichiara}, {Gropp}, {Kennea}, {Kuin},
  {Lien}, {Marshall}, {Tohuvavohu}, {Williams}, \& {Neil Gehrels Swift
  Observatory Team}}]{GCN32632}
{Dichiara}, S., {Gropp}, J.~D., {Kennea}, J.~A., {et~al.} 2022, GRB Coordinates
  Network, 32632, 1

\bibitem[{{Feroz} {et~al.}(2009){Feroz}, {Hobson}, \& {Bridges}}]{Feroz09}
{Feroz}, F., {Hobson}, M.~P., \& {Bridges}, M. 2009, \mnras, 398, 1601,
  \dodoi{10.1111/j.1365-2966.2009.14548.x}

\bibitem[{{Ferro} {et~al.}(2022){Ferro}, {Brivio}, {D'Avanzo}, {Piranomonte},
  {Lorenzi}, {Mainella}, \& {CIBO Collaboration}}]{GCN.32804}
{Ferro}, M., {Brivio}, R., {D'Avanzo}, P., {et~al.} 2022, GRB Coordinates
  Network, 32804, 1

\bibitem[{{Filippenko}(1997)}]{Filippenko1997}
{Filippenko}, A.~V. 1997, \araa, 35, 309,
  \dodoi{10.1146/annurev.astro.35.1.309}

\bibitem[{{Flewelling} {et~al.}(2020){Flewelling}, {Magnier}, {Chambers},
  {Heasley}, {Holmberg}, {Huber}, {Sweeney}, {Waters}, {Calamida}, {Casertano},
  {Chen}, {Farrow}, {Hasinger}, {Henderson}, {Long}, {Metcalfe}, {Narayan},
  {Nieto-Santisteban}, {Norberg}, {Rest}, {Saglia}, {Szalay}, {Thakar},
  {Tonry}, {Valenti}, {Werner}, {White}, {Denneau}, {Draper}, {Hodapp},
  {Jedicke}, {Kaiser}, {Kudritzki}, {Price}, {Wainscoat}, {Chastel}, {McLean},
  {Postman}, \& {Shiao}}]{PS1}
{Flewelling}, H.~A., {Magnier}, E.~A., {Chambers}, K.~C., {et~al.} 2020, \apjs,
  251, 7, \dodoi{10.3847/1538-4365/abb82d}

\bibitem[{{Foreman-Mackey} {et~al.}(2013){Foreman-Mackey}, {Hogg}, {Lang}, \&
  {Goodman}}]{emcee}
{Foreman-Mackey}, D., {Hogg}, D.~W., {Lang}, D., \& {Goodman}, J. 2013, \pasp,
  125, 306, \dodoi{10.1086/670067}

\bibitem[{{Frederiks} {et~al.}(2023){Frederiks}, {Svinkin}, {Lysenko},
  {Molkov}, {Tsvetkova}, {Ulanov}, {Ridnaia}, {Lutovinov}, {Lapshov},
  {Tkachenko}, \& {Levin}}]{Frederiks2023}
{Frederiks}, D., {Svinkin}, D., {Lysenko}, A.~L., {et~al.} 2023, arXiv
  e-prints, arXiv:2302.13383, \dodoi{10.48550/arXiv.2302.13383}

\bibitem[{{Fulton} {et~al.}(2023){Fulton}, {Smartt}, {Rhodes}, {Huber},
  {Villar}, {Moore}, {Srivastav}, {Schultz}, {Chambers}, {Izzo}, {Hjorth},
  {Chen}, {Nicholl}, {Foley}, {Rest}, {Smith}, {Young}, {Sim}, {Bright},
  {Zenati}, {de Boer}, {Bulger}, {Fairlamb}, {Gao}, {Lin}, {Lowe}, {Magnier},
  {Smith}, {Wainscoat}, {Coulter}, {Jones}, {Kilpatrick}, {McGill},
  {Ramirez-Ruiz}, {Lee}, {Narayan}, {Ramakrishnan}, {Ridden-Harper}, {Singh},
  {Wang}, {Kong}, {Ngeow}, {Pan}, {Yang}, {Davis}, {Piro}, {Rojas-Bravo},
  {Sommer}, \& {Yadavalli}}]{Fulton2023}
{Fulton}, M.~D., {Smartt}, S.~J., {Rhodes}, L., {et~al.} 2023, arXiv e-prints,
  arXiv:2301.11170, \dodoi{10.48550/arXiv.2301.11170}

\bibitem[{Gehrels {et~al.}(2004)Gehrels, Chincarini, Giommi, Mason, Nousek,
  Wells, White, Barthelmy, Burrows, Cominsky, Hurley, Marshall, Meszaros,
  Roming, Angelini, Barbier, Belloni, Campana, Caraveo, Chester, Citterio,
  Cline, Cropper, Cummings, Dean, Feigelson, Fenimore, Frail, Fruchter,
  Garmire, Gendreau, Ghisellini, Greiner, Hill, Hunsberger, Krimm, Kulkarni,
  Kumar, Lebrun, Lloyd-Ronning, Markwardt, Mattson, Mushotzky, Norris, Osborne,
  Paczynski, Palmer, Park, Parsons, Paul, Rees, Reynolds, Rhoads, Sasseen,
  Schaefer, Short, Smale, Smith, Stella, Tagliaferri, Takahashi, Tashiro,
  Townsley, Tueller, Turner, Vietri, Voges, Ward, Willingale, Zerbi, \&
  Zhang}]{gehrels_2004}
Gehrels, N., Chincarini, G., Giommi, P., {et~al.} 2004, \apj, 611, 1005,
  \dodoi{10.1086/422091}

\bibitem[{{Groot} {et~al.}(2022){Groot}, {Vreeswijk}, {Ter Horst}, {Bloemen},
  {Jonker}, {de Wet}, {Malesani}, {Pieterse}, \& {BlackGEM
  Consortium}}]{GCN.32678}
{Groot}, P.~J., {Vreeswijk}, P.~M., {Ter Horst}, R., {et~al.} 2022, GRB
  Coordinates Network, 32678, 1

\bibitem[{{Gupta} {et~al.}(2022){Gupta}, {Ror}, {Pandey}, {Aryan}, {Ghosh},
  {\%Dimple}, \& {Misra}}]{GCN.32811}
{Gupta}, R., {Ror}, A.~K., {Pandey}, S.~B., {et~al.} 2022, GRB Coordinates
  Network, 32811, 1

\bibitem[{{Hjorth}(2013)}]{hjorth2013}
{Hjorth}, J. 2013, Philosophical Transactions of the Royal Society of London
  Series A, 371, 20120275, \dodoi{10.1098/rsta.2012.0275}

\bibitem[{{Huber} {et~al.}(2022){Huber}, {Schultz}, {Chambers}, {Smith},
  {Fulton}, {Smartt}, {Chen}, {Nicholl}, {Young}, {Shingles}, {Srivastav},
  {Sim}, {de Boer}, {Bulger}, {Fairlamb}, {Lin}, {Lowe}, {Magnier},
  {Wainscoat}, {Gao}, {Stubbs}, \& {Rest}}]{GCN.32758}
{Huber}, M., {Schultz}, A., {Chambers}, K.~C., {et~al.} 2022, GRB Coordinates
  Network, 32758, 1

\bibitem[{{Im} {et~al.}(2022){Im}, {Paek}, {Lim}, {Choi}, {Kim}, {Sung}, \&
  {Urata}}]{GCN.32803}
{Im}, M., {Paek}, G. S.~H., {Lim}, G., {et~al.} 2022, GRB Coordinates Network,
  32803, 1

\bibitem[{{Iwamoto} {et~al.}(1998){Iwamoto}, {Mazzali}, {Nomoto}, {Umeda},
  {Nakamura}, {Patat}, {Danziger}, {Young}, {Suzuki}, {Shigeyama},
  {Augusteijn}, {Doublier}, {Gonzalez}, {Boehnhardt}, {Brewer}, {Hainaut},
  {Lidman}, {Leibundgut}, {Cappellaro}, {Turatto}, {Galama}, {Vreeswijk},
  {Kouveliotou}, {van Paradijs}, {Pian}, {Palazzi}, \& {Frontera}}]{Iwamoto98}
{Iwamoto}, K., {Mazzali}, P.~A., {Nomoto}, K., {et~al.} 1998, \nat, 395, 672,
  \dodoi{10.1038/27155}

\bibitem[{{Izzo} {et~al.}(2022){Izzo}, {Saccardi}, {Fynbo}, {Palmerio},
  {Malesani}, {Agui Fernandez}, {Kann}, {Melandri}, {Vergani}, {Wiersema}, \&
  {Stargate Consortium}}]{GCN32765hostz}
{Izzo}, L., {Saccardi}, A., {Fynbo}, J.~P.~U., {et~al.} 2022, GRB Coordinates
  Network, 32765, 1

\bibitem[{{Jakobsson} {et~al.}(2006){Jakobsson}, {Levan}, {Fynbo}, {Priddey},
  {Hjorth}, {Tanvir}, {Watson}, {Jensen}, {Sollerman}, {Natarajan},
  {Gorosabel}, {Castro Cer{\'o}n}, {Pedersen}, {Pursimo}, {{\'A}rnad{\'o}ttir},
  {Castro-Tirado}, {Davis}, {Deeg}, {Fiuza}, {Mikolaitis}, \&
  {Sousa}}]{Jakobsson+2006}
{Jakobsson}, P., {Levan}, A., {Fynbo}, J.~P.~U., {et~al.} 2006, \aap, 447, 897,
  \dodoi{10.1051/0004-6361:20054287}

\bibitem[{{Jester} {et~al.}(2005){Jester}, {Schneider}, {Richards}, {Green},
  {Schmidt}, {Hall}, {Strauss}, {Vanden Berk}, {Stoughton}, {Gunn},
  {Brinkmann}, {Kent}, {Smith}, {Tucker}, \& {Yanny}}]{Jester2005}
{Jester}, S., {Schneider}, D.~P., {Richards}, G.~T., {et~al.} 2005, \aj, 130,
  873, \dodoi{10.1086/432466}

\bibitem[{{Kann} {et~al.}(2007){Kann}, {Masetti}, \& {Klose}}]{Kann2007}
{Kann}, D.~A., {Masetti}, N., \& {Klose}, S. 2007, \aj, 133, 1187,
  \dodoi{10.1086/511066}

\bibitem[{{Kann} {et~al.}(2023){Kann}, {Agayeva}, {Aivazyan}, {Alishov},
  {Andrade}, {Antier}, {Baransky}, {Bendjoya}, {Benkhaldoun}, {Beradze},
  {Berezin}, {Bo{\"e}r}, {Broens}, {Brunier}, {Bulla}, {Burkhonov}, {Burns},
  {Chen}, {Chen}, {Conti}, {Coughlin}, {Cui}, {Daigne}, {Delaveau},
  {Devillepoix}, {Dietrich}, {Dornic}, {Dubois}, {Ducoin}, {Durand}, {Duverne},
  {Eggenstein}, {Ehgamberdiev}, {Fouad}, {Freeberg}, {Froebrich}, {Ge},
  {Gervasoni}, {Godunova}, {Gokuldass}, {Gurbanov}, {Han}, {Hasanov}, {Hello},
  {Hussenot-Desenonges}, {Inasaridze}, {Iskandar}, {Ismailov}, {Janati}, {Jegou
  du Laz}, {Jia}, {Karpov}, {Kaeouach}, {Kiendrebeogo}, {Klotz}, {Kneip},
  {Kochiashvili}, {Kunert}, {Lekic}, {Leonini}, {Li}, {Li}, {Li}, {Liao},
  {Logie}, {Lu}, {Mao}, {Marchais}, {M{\'e}nard}, {Morris}, {Natsvlishvili},
  {Nedora}, {Noonan}, {Noysena}, {Orange}, {Pang}, {Peng}, {Pellouin},
  {Peloton}, {Pradier}, {Pyshna}, {Rajabo}, {Rau}, {Rinner}, {Rivet},
  {Romanov}, {Rosi}, {Rupchandani}, {Serrau}, {Shokry}, {Simon}, {Smith},
  {Sokoliuk}, {Soliman}, {Song}, {Takey}, {Tillayev}, {Tinjaca Ramirez}, {Tosta
  e Melo}, {Turpin}, {de Ugarte Postigo}, {Vanaverbeke}, {Vasylenko}, {Vernet},
  {Vidadi}, {Wang}, {Wang}, {Wang}, {Wang}, {Xiong}, {Xu}, {Xue}, {Zeng},
  {Zhang}, {Zhao}, \& {Zhao}}]{Kann+2023}
{Kann}, D.~A., {Agayeva}, S., {Aivazyan}, V., {et~al.} 2023, arXiv e-prints,
  arXiv:2302.06225, \dodoi{10.48550/arXiv.2302.06225}

\bibitem[{{Kim} {et~al.}(2022){Kim}, {Krugov}, {Pozanenko}, {Aimuratov},
  {Belkin}, {Pankov}, \& {IKI FuN}}]{GCN.32670}
{Kim}, V., {Krugov}, M., {Pozanenko}, A., {et~al.} 2022, GRB Coordinates
  Network, 32670, 1

\bibitem[{{Klose} {et~al.}(2019){Klose}, {Schmidl}, {Kann}, {Nicuesa
  Guelbenzu}, {Schulze}, {Greiner}, {Olivares E.}, {Kr{\"u}hler}, {Schady},
  {Afonso}, {Filgas}, {Fynbo}, {Rau}, {Rossi}, {Takats}, {Tanga}, {Updike}, \&
  {Varela}}]{Klose2019}
{Klose}, S., {Schmidl}, S., {Kann}, D.~A., {et~al.} 2019, \aap, 622, A138,
  \dodoi{10.1051/0004-6361/201832728}

\bibitem[{{Kumar} {et~al.}(2022{\natexlab{a}}){Kumar}, {Swain}, {Waratkar},
  {Angail}, {Bhalerao}, {Anupama}, {Barway}, \& {GIT Team}}]{GCN.32662}
{Kumar}, H., {Swain}, V., {Waratkar}, G., {et~al.} 2022{\natexlab{a}}, GRB
  Coordinates Network, 32662, 1

\bibitem[{{Kumar} {et~al.}(2022{\natexlab{b}}){Kumar}, {Bhalerao}, {Anupama},
  {Barway}, {Basu}, {Deshmukh}, {De}, {Dutta}, {Fremling}, {Iyer}, {Jassani},
  {Joharle}, {Karambelkar}, {Khandagale}, {Krishna}, {Kulkarni}, {Mate},
  {Patil}, {Phanindra}, {Samantaray}, {Sharma}, {Sharma}, {Shenoy}, {Singh},
  {Srivastava}, {Swain}, {Waratkar}, {Angchuk}, {Dorjay}, {Dorjai}, {Gyalson},
  {Jorphail}, {Mahay}, {Norbu}, {Sharma}, {Stanzin}, {Stanzin}, \&
  {Stanzin}}]{2022AJ....164...90K}
{Kumar}, H., {Bhalerao}, V., {Anupama}, G.~C., {et~al.} 2022{\natexlab{b}},
  \aj, 164, 90, \dodoi{10.3847/1538-3881/ac7bea}

\bibitem[{{Kumar} {et~al.}(2022{\natexlab{c}}){Kumar}, {Bhalerao}, {Anupama},
  {Barway}, {Coughlin}, {De}, {Deshmukh}, {Dutta}, {Goldstein}, {Jassani},
  {Joharle}, {Karambelker}, {Khandagale}, {Kumar}, {Saraogi}, {Sharma},
  {Shenoy}, {singer}, {Singh}, \& {Waratkar}}]{2022MNRAS.516.4517K}
---. 2022{\natexlab{c}}, \mnras, 516, 4517, \dodoi{10.1093/mnras/stac2516}

\bibitem[{{Labrie} {et~al.}(2019){Labrie}, {Anderson}, {C{\'a}rdenes},
  {Simpson}, \& {Turner}}]{Labrie2019}
{Labrie}, K., {Anderson}, K., {C{\'a}rdenes}, R., {Simpson}, C., \& {Turner},
  J. E.~H. 2019, in Astronomical Society of the Pacific Conference Series, Vol.
  523, Astronomical Data Analysis Software and Systems XXVII, ed. P.~J.
  {Teuben}, M.~W. {Pound}, B.~A. {Thomas}, \& E.~M. {Warner}, 321

\bibitem[{{Laskar} {et~al.}(2022){Laskar}, {Alexander}, {Ayache}, {Berger},
  {Chornock}, {van Eerten}, {Fong}, {Margutti}, {Mundell}, \&
  {Schady}}]{GCN.32757}
{Laskar}, T., {Alexander}, K.~D., {Ayache}, E., {et~al.} 2022, GRB Coordinates
  Network, 32757, 1

\bibitem[{{Laskar} {et~al.}(2023){Laskar}, {Alexander}, {Margutti},
  {Eftekhari}, {Chornock}, {Berger}, {Cendes}, {Duerr}, {Perley}, {Edvige
  Ravasio}, {Yamazaki}, {Ayache}, {Barclay}, {Barniol Duran}, {Bhandari},
  {Brethauer}, {Christy}, {Coppejans}, {Duffell}, {Fong}, {Gomboc}, {Guidorzi},
  {Kennea}, {Kobayashi}, {Levan}, {Lobanov}, {Metzger}, {Ros}, {Schroeder}, \&
  {Williams}}]{Laskar+2023}
{Laskar}, T., {Alexander}, K.~D., {Margutti}, R., {et~al.} 2023, arXiv
  e-prints, arXiv:2302.04388, \dodoi{10.48550/arXiv.2302.04388}

\bibitem[{{Lesage} {et~al.}(2022){Lesage}, {Veres}, {Roberts}, {Burns},
  {Bissaldi}, \& {Fermi GBM Team}}]{GBMGCN}
{Lesage}, S., {Veres}, P., {Roberts}, O.~J., {et~al.} 2022, GRB Coordinates
  Network, 32642, 1

\bibitem[{{Levan} {et~al.}(2023){Levan}, {Lamb}, {Schneider}, {Hjorth},
  {Zafar}, {de Ugarte Postigo}, {Sargent}, {Mullally}, {Izzo}, {D'Avanzo},
  {Burns}, {Ag{\"u}{\'\i} Fern{\'a}ndez}, {Barclay}, {Bernardini},
  {Bhirombhakdi}, {Bremer}, {Brivio}, {Campana}, {Chrimes}, {D'Elia}, {De
  Pasquale}, {Ferro}, {Fong}, {Fruchter}, {Fynbo}, {Gaspari}, {Gompertz},
  {Hartmann}, {Hedges}, {Heintz}, {Hotokezaka}, {Jakobsson}, {Kann}, {Kennea},
  {Laskar}, {Le Floc'h}, {Malesani}, {Melandri}, {Metzger}, {Oates}, {Pian},
  {Piranomonte}, {Pugliese}, {Racusin}, {Rastinejad}, {Ravasio}, {Rossi},
  {Saccardi}, {Salvaterra}, {Sbarufatti}, {Starling}, {Tanvir}, {Th{\"o}ne},
  {Vergani}, {Watson}, {Wiersema}, \& {Xu}}]{Levan2023}
{Levan}, A.~J., {Lamb}, G.~P., {Schneider}, B., {et~al.} 2023, arXiv e-prints,
  arXiv:2302.07761.
\newblock \doarXiv{2302.07761}

\bibitem[{{Lyman} {et~al.}(2014){Lyman}, {Bersier}, \& {James}}]{Lyman2014}
{Lyman}, J.~D., {Bersier}, D., \& {James}, P.~A. 2014, \mnras, 437, 3848,
  \dodoi{10.1093/mnras/stt2187}

\bibitem[{{Lyman} {et~al.}(2016){Lyman}, {Bersier}, {James}, {Mazzali},
  {Eldridge}, {Fraser}, \& {Pian}}]{Lyman2016}
{Lyman}, J.~D., {Bersier}, D., {James}, P.~A., {et~al.} 2016, \mnras, 457, 328,
  \dodoi{10.1093/mnras/stv2983}

\bibitem[{{Malesani} {et~al.}(2023){Malesani}, {Levan}, {Izzo}, {de Ugarte
  Postigo}, {Ghirlanda}, {Heintz}, {Kann}, {Lamb}, {Palmerio}, {Salafia},
  {Salvaterra}, {Tanvir}, {Ag{\"u}{\'\i} Fern{\'a}ndez}, {Campana}, {Chrimes},
  {D'Avanzo}, {D'Elia}, {Della Valle}, {De Pasquale}, {Fynbo}, {Gaspari},
  {Gompertz}, {Hartmann}, {Hjorth}, {Jakobsson}, {Palazzi}, {Pian}, {Pugliese},
  {Ravasio}, {Rossi}, {Saccardi}, {Schady}, {Schneider}, {Sollerman},
  {Starling}, {Th{\"o}ne}, {van der Horst}, {Vergani}, {Watson}, {Wiersema},
  {Xu}, \& {Zafar}}]{Malesani2023}
{Malesani}, D.~B., {Levan}, A.~J., {Izzo}, L., {et~al.} 2023, arXiv e-prints,
  arXiv:2302.07891, \dodoi{10.48550/arXiv.2302.07891}

\bibitem[{{McCully} \& {Tewes}(2019)}]{2019ascl.soft07032M}
{McCully}, C., \& {Tewes}, M. 2019, {Astro-SCRAPPY: Speedy Cosmic Ray
  Annihilation Package in Python}, Astrophysics Source Code Library.
\newblock \doeprint{1907.032}

\bibitem[{{Meegan} {et~al.}(2009){Meegan}, {Lichti}, {Bhat}, {Bissaldi},
  {Briggs}, {Connaughton}, {Diehl}, {Fishman}, {Greiner}, {Hoover}, {van der
  Horst}, {von Kienlin}, {Kippen}, {Kouveliotou}, {McBreen}, {Paciesas},
  {Preece}, {Steinle}, {Wallace}, {Wilson}, \& {Wilson-Hodge}}]{Meegan+2009}
{Meegan}, C., {Lichti}, G., {Bhat}, P.~N., {et~al.} 2009, \apj, 702, 791,
  \dodoi{10.1088/0004-637X/702/1/791}

\bibitem[{{Melandri} {et~al.}(2019){Melandri}, {Malesani}, {Izzo}, {Japelj},
  {Vergani}, {Schady}, {Sagu{\'e}s Carracedo}, {de Ugarte Postigo}, {Anderson},
  {Barbarino}, {Bolmer}, {Breeveld}, {Calissendorff}, {Campana}, {Cano},
  {Carini}, {Covino}, {D'Avanzo}, {D'Elia}, {della Valle}, {De Pasquale},
  {Fynbo}, {Gromadzki}, {Hammer}, {Hartmann}, {Heintz}, {Inserra}, {Jakobsson},
  {Kann}, {Kotilainen}, {Maguire}, {Masetti}, {Nicholl}, {Olivares E},
  {Pugliese}, {Rossi}, {Salvaterra}, {Sollerman}, {Stone}, {Tagliaferri},
  {Tomasella}, {Th{\"o}ne}, {Xu}, \& {Young}}]{Melandri2019}
{Melandri}, A., {Malesani}, D.~B., {Izzo}, L., {et~al.} 2019, \mnras, 490,
  5366, \dodoi{10.1093/mnras/stz2900}

\bibitem[{{M{\'e}sz{\'a}ros} \& {Rees}(1997)}]{Meszaros1997}
{M{\'e}sz{\'a}ros}, P., \& {Rees}, M.~J. 1997, \apj, 476, 232,
  \dodoi{10.1086/303625}

\bibitem[{{Oates} {et~al.}(2009){Oates}, {Page}, {Schady}, {de Pasquale},
  {Koch}, {Breeveld}, {Brown}, {Chester}, {Holland}, {Hoversten}, {Kuin},
  {Marshall}, {Roming}, {Still}, {vanden Berk}, {Zane}, \&
  {Nousek}}]{Oates2009}
{Oates}, S.~R., {Page}, M.~J., {Schady}, P., {et~al.} 2009, \mnras, 395, 490,
  \dodoi{10.1111/j.1365-2966.2009.14544.x}

\bibitem[{{O'Connor} {et~al.}(2022{\natexlab{a}}){O'Connor}, {Cenko}, {Troja},
  {Dichiara}, {Kutyrev}, {Veilleux}, \& {Durbak}}]{GCN32739}
{O'Connor}, B., {Cenko}, S.~B., {Troja}, E., {et~al.} 2022{\natexlab{a}}, GRB
  Coordinates Network, 32739, 1

\bibitem[{{O'Connor} {et~al.}(2022{\natexlab{b}}){O'Connor}, {Cenko}, {Troja},
  {Dichiara}, {Kutyrev}, {Veilleux}, \& {Durbak}}]{GCN32799}
---. 2022{\natexlab{b}}, GRB Coordinates Network, 32799, 1

\bibitem[{{O'Connor} {et~al.}(2023){O'Connor}, {Troja}, {Ryan}, {Beniamini},
  {van Eerten}, {Granot}, {Dichiara}, {Ricci}, {Lipunov}, {Gillanders}, {Gill},
  {Moss}, {Anand}, {Andreoni}, {Becerra}, {Buckley}, {Butler}, {Cenko},
  {Chasovnikov}, {Durbak}, {Francile}, {Hammerstein}, {van der Horst},
  {Kasliwal}, {Kouveliotou}, {Kutyrev}, {Lee}, {Srinivasaragavan}, {Topolev},
  {Watson}, {Yang}, \& {Zhirkov}}]{OConnor+2023}
{O'Connor}, B., {Troja}, E., {Ryan}, G., {et~al.} 2023, arXiv e-prints,
  arXiv:2302.07906, \dodoi{10.48550/arXiv.2302.07906}

\bibitem[{{Pellegrin} {et~al.}(2022){Pellegrin}, {Rumstay}, \&
  {Hartmann}}]{GCN.32852}
{Pellegrin}, K., {Rumstay}, K., \& {Hartmann}, D. 2022, GRB Coordinates
  Network, 32852, 1

\bibitem[{{Postigo} {et~al.}(2022){Postigo}, {Izzo}, {Thoene}, {Fynbo}, {Kann},
  {Fernandez}, \& {Tanvir}}]{TNS}
{Postigo}, A.~D.~U., {Izzo}, L., {Thoene}, C.~C., {et~al.} 2022, Transient Name
  Server Classification Report, 2022-3047, 1

\bibitem[{{Rajabov} {et~al.}(2022){Rajabov}, {Sadibekova}, {Tillayev},
  {Rinner}, {Benkhaldoun}, {Wang}, {Zhu}, {Zeng}, {Wang}, {Iskandar}, {Fouad},
  {Shokry}, {Takey}, {Soliman}, {Hello}, {Hussenot}, {Boer}, {de Ugarte
  Postigo}, {Antier}, {Kann}, {Burns}, {Simon}, {Baransky}, {Abe}, {Bendjoya},
  {Rivet}, {Vernet}, {Brunier}, {Inasaridze}, {Natsvlishvili}, {Kochiashvili},
  {Beradze}, {Aivazyan}, {Kapanadze}, {Burkhonov}, {Ducoin}, {Ehgamberdiev},
  {Klotz}, {Tosta E. Melo}, \& {GRANDMA Collaboration}}]{GCN.32795}
{Rajabov}, Y., {Sadibekova}, T., {Tillayev}, Y., {et~al.} 2022, GRB Coordinates
  Network, 32795, 1

\bibitem[{{Rastinejad} \& {Fong}(2022)}]{GCN32749}
{Rastinejad}, J., \& {Fong}, W. 2022, GRB Coordinates Network, 32749, 1

\bibitem[{{Rossi} {et~al.}(2022){Rossi}, {Maiorano}, {Malesani}, {CIBO
  Collaboration}, {Cusano}, \& {Paris}}]{GCN.32809}
{Rossi}, A., {Maiorano}, E., {Malesani}, D.~B., {et~al.} 2022, GRB Coordinates
  Network, 32809, 1

\bibitem[{{Sari} {et~al.}(1998){Sari}, {Piran}, \& {Narayan}}]{Sari1998}
{Sari}, R., {Piran}, T., \& {Narayan}, R. 1998, \apjl, 497, L17,
  \dodoi{10.1086/311269}

\bibitem[{{Schlafly} \& {Finkbeiner}(2011)}]{Schlafly2011}
{Schlafly}, E.~F., \& {Finkbeiner}, D.~P. 2011, \apj, 737, 103,
  \dodoi{10.1088/0004-637X/737/2/103}

\bibitem[{{Schneider} {et~al.}(2022){Schneider}, {Adami}, {Le Floc'h},
  {Turpin}, {G{\"o}tz}, {Vergani}, {Saccardi}, {Basa}, {Le Van Suu}, \& {a
  larger Collaboration}}]{GCN.32753}
{Schneider}, B., {Adami}, C., {Le Floc'h}, E., {et~al.} 2022, GRB Coordinates
  Network, 32753, 1

\bibitem[{{Shrestha} {et~al.}(2022){Shrestha}, {Bostroem}, {Sand}, {Alexander},
  {Andrews}, {Pearson}, {Hosseinzadeh}, {Smith}, {Howell}, {McCully},
  {Newsome}, {Padilla Gonzalez}, {Pellegrino}, {Terreran}, {Farah}, \& {Global
  Supernova Project Collaboration}}]{GCN.32771}
{Shrestha}, M., {Bostroem}, K., {Sand}, D., {et~al.} 2022, GRB Coordinates
  Network, 32771, 1

\bibitem[{{Shrestha} {et~al.}(2023){Shrestha}, {Sand}, {Alexander}, {Bostroem},
  {Hosseinzadeh}, {Pearson}, {Aghakhanloo}, {Vink{\'o}}, {Andrews}, {Jencson},
  {Lundquist}, {Wyatt}, {Howell}, {McCully}, {Padilla Gonzalez}, {Pellegrino},
  {Terreran}, {Hiramatsu}, {Newsome}, {Farah}, {Jha}, {Smith}, {Wheeler},
  {Mart{\'\i}nez-V{\'a}zquez}, {Carballo-Bello}, {Drlica-Wagner}, {James},
  {Mutlu-Pakdil}, {Stringfellow}, {Sakowska}, {No{\"e}l}, {Bom}, \&
  {Kuehn}}]{Shrestha+2023}
{Shrestha}, M., {Sand}, D.~J., {Alexander}, K.~D., {et~al.} 2023, arXiv
  e-prints, arXiv:2302.03829, \dodoi{10.48550/arXiv.2302.03829}

\bibitem[{{Soderberg} {et~al.}(2006){Soderberg}, {Nakar}, {Berger}, \&
  {Kulkarni}}]{Soderberg+2006}
{Soderberg}, A.~M., {Nakar}, E., {Berger}, E., \& {Kulkarni}, S.~R. 2006, \apj,
  638, 930, \dodoi{10.1086/499121}

\bibitem[{{Sollerman} {et~al.}(2000){Sollerman}, {Kozma}, {Fransson},
  {Leibundgut}, {Lundqvist}, {Ryde}, \& {Woudt}}]{Sollerman2000}
{Sollerman}, J., {Kozma}, C., {Fransson}, C., {et~al.} 2000, \apjl, 537, L127,
  \dodoi{10.1086/312763}

\bibitem[{{Stritzinger} {et~al.}(2002){Stritzinger}, {Hamuy}, {Suntzeff},
  {Smith}, {Candia}, {Phillips}, {Maza}, {Antezana}, {Wischnjewsky},
  {Strolger}, {Stubbs}, {Becker}, {Rubenstein}, {Galaz}, \&
  {Gonzalez}}]{Stritzinger2002}
{Stritzinger}, M.~D., {Hamuy}, M., {Suntzeff}, N.~B., {et~al.} 2002, in
  American Astronomical Society Meeting Abstracts, Vol. 200, American
  Astronomical Society Meeting Abstracts \#200, 95.03

\bibitem[{{Taddia} {et~al.}(2018){Taddia}, {Stritzinger}, {Bersten}, {Baron},
  {Burns}, {Contreras}, {Holmbo}, {Hsiao}, {Morrell}, {Phillips}, {Sollerman},
  \& {Suntzeff}}]{Taddia2018a}
{Taddia}, F., {Stritzinger}, M.~D., {Bersten}, M., {et~al.} 2018, \aap, 609,
  A136, \dodoi{10.1051/0004-6361/201730844}

\bibitem[{{Taddia} {et~al.}(2019){Taddia}, {Sollerman}, {Fremling},
  {Barbarino}, {Karamehmetoglu}, {Arcavi}, {Cenko}, {Filippenko}, {Gal-Yam},
  {Hiramatsu}, {Hosseinzadeh}, {Howell}, {Kulkarni}, {Laher}, {Lunnan},
  {Masci}, {Nugent}, {Nyholm}, {Perley}, {Quimby}, \& {Silverman}}]{Taddia2018}
{Taddia}, F., {Sollerman}, J., {Fremling}, C., {et~al.} 2019, \aap, 621, A71,
  \dodoi{10.1051/0004-6361/201834429}

\bibitem[{{Tiengo} {et~al.}(2023){Tiengo}, {Pintore}, {Vaia}, {Filippi},
  {Sacchi}, {Esposito}, {Rigoselli}, {Mereghetti}, {Salvaterra}, {Siljeg},
  {Bracco}, {Bosnjak}, {Jelic}, \& {Campana}}]{Tiengo2023}
{Tiengo}, A., {Pintore}, F., {Vaia}, B., {et~al.} 2023, arXiv e-prints,
  arXiv:2302.11518, \dodoi{10.48550/arXiv.2302.11518}

\bibitem[{{Toy} {et~al.}(2016){Toy}, {Cenko}, {Silverman}, {Butler},
  {Cucchiara}, {Watson}, {Bersier}, {Perley}, {Margutti}, {Bellm}, {Bloom},
  {Cao}, {Capone}, {Clubb}, {Corsi}, {De Cia}, {de Diego}, {Filippenko}, {Fox},
  {Gal-Yam}, {Gehrels}, {Georgiev}, {Gonz{\'a}lez}, {Kasliwal}, {Kelly},
  {Kulkarni}, {Kutyrev}, {Lee}, {Prochaska}, {Ramirez-Ruiz}, {Richer},
  {Rom{\'a}n-Z{\'u}{\~n}iga}, {Singer}, {Stern}, {Troja}, \&
  {Veilleux}}]{Toy2016}
{Toy}, V.~L., {Cenko}, S.~B., {Silverman}, J.~M., {et~al.} 2016, \apj, 818, 79,
  \dodoi{10.3847/0004-637X/818/1/79}

\bibitem[{{Trotta}(2008)}]{Trota08}
{Trotta}, R. 2008, Contemporary Physics, 49, 71,
  \dodoi{10.1080/00107510802066753}

\bibitem[{{Valenti} {et~al.}(2008){Valenti}, {Benetti}, {Cappellaro}, {Patat},
  {Mazzali}, {Turatto}, {Hurley}, {Maeda}, {Gal-Yam}, {Foley}, {Filippenko},
  {Pastorello}, {Challis}, {Frontera}, {Harutyunyan}, {Iye}, {Kawabata},
  {Kirshner}, {Li}, {Lipkin}, {Matheson}, {Nomoto}, {Ofek}, {Ohyama}, {Pian},
  {Poznanski}, {Salvo}, {Sauer}, {Schmidt}, {Soderberg}, \&
  {Zampieri}}]{Valenti2008}
{Valenti}, S., {Benetti}, S., {Cappellaro}, E., {et~al.} 2008, \mnras, 383,
  1485, \dodoi{10.1111/j.1365-2966.2007.12647.x}

\bibitem[{{Vinko} {et~al.}(2022){Vinko}, {Bodi}, {Pal}, {Kriskovics},
  {Szakats}, \& {Vida}}]{GCN.32709}
{Vinko}, J., {Bodi}, A., {Pal}, A., {et~al.} 2022, GRB Coordinates Network,
  32709, 1

\bibitem[{{Williams} {et~al.}(2023){Williams}, {Kennea}, {Dichiara},
  {Kobayashi}, {Iwakiri}, {Beardmore}, {Evans}, {Heinz}, {Lien}, {Oates},
  {Negoro}, {Cenko}, {Buisson}, {Hartmann}, {Jaisawal}, {Kuin}, {Lesage},
  {Page}, {Parsotan}, {Pasham}, {Sbarufatti}, {Siegel}, {Sugita}, {Younes},
  {Ambrosi}, {Arzoumanian}, {Bernardini}, {Campana}, {Capalbi}, {Caputo},
  {D'Ai}, {D'Avanzo}, {D'Elia}, {De Pasquale}, {Eyles-Ferris}, {Ferrara},
  {Gendreau}, {Gropp}, {Kawai}, {Klingler}, {Laha}, {Melandri}, {Mihara},
  {Moss}, {O'Brien}, {Osborne}, {Palmer}, {Perri}, {Serino}, {Sonbas},
  {Stamatikos}, {Starling}, {Tagliaferri}, {Tohuvavohu}, {Zane}, \&
  {Ziaeepour}}]{Williams+2023}
{Williams}, M.~A., {Kennea}, J.~A., {Dichiara}, S., {et~al.} 2023, arXiv
  e-prints, arXiv:2302.03642, \dodoi{10.48550/arXiv.2302.03642}

\bibitem[{{Woosley} \& {Bloom}(2006)}]{Woosley2006}
{Woosley}, S.~E., \& {Bloom}, J.~S. 2006, \araa, 44, 507,
  \dodoi{10.1146/annurev.astro.43.072103.150558}

\bibitem[{{Yang} \& {Sollerman}(2023)}]{Yang+2023}
{Yang}, S., \& {Sollerman}, J. 2023, arXiv e-prints, arXiv:2302.02082,
  \dodoi{10.48550/arXiv.2302.02082}

\end{thebibliography}
\bibliographystyle{aasjournal}



\end{document}